\documentclass[seceqn]{elsart}

\renewcommand{\thesection}{\arabic{section}.}

\usepackage{graphicx}
\usepackage{bm}

\begin{document}

\begin{frontmatter}



\title{Photon splitting in atomic fields}


\author{R.N.~Lee, A.L.~Maslennikov, A.I.~Milstein,}
\author{ V.M.~Strakhovenko, and Yu.A.~Tikhonov}

\address{G.I. Budker Institute of Nuclear Physics,\\
630090 Novosibirsk, Russia}

\begin{abstract}
Photon splitting due to vacuum polarization in the electric field
of an atom is considered. We survey different theoretical
approaches to the description of this nonlinear QED process and
several attempts of its experimental observation. We present the
results of the lowest-order perturbation theory as well as those
obtained within the quasiclassical approximation being exact in
the external field strength. The experiment where photon splitting
was really observed for the first time is discussed in details.
The results of this experiment are compared with recent
theoretical estimations.
\end{abstract}

\begin{keyword}
photon splitting \sep nonlinear QED processes \sep external field
\sep Green function
\PACS 12.20.Ds \sep 12.20.Fv
\end{keyword}
\end{frontmatter}
\tableofcontents

\begin{flushright}
{\it To the memory of our\\ friend and colleague \\Guram
Kezerashvili}
\end{flushright}

\section{Introduction}

Virtual creation and annihilation of electron-positron pairs is
known to induce a self-action of an electromagnetic field, which
results in such effects as coherent photon scattering and photon
splitting in an external field, and photon-photon scattering.
Among these processes the latter was the first one subjected to
the exhaustive theoretical investigation \cite{DeTollis64} (see
also \cite{CTP} and references therein). However, this process was
never observed experimentally. At present, coherent photon
scattering in the electric field of atoms (Delbr\"uck scattering)
is investigated in detail both theoretically and experimentally
\cite{PM75,MShu94,Akh98}. The amplitudes and cross sections of
this process were calculated in the lowest order in $Z\alpha$
(Born approximation) for arbitrary photon energy $\omega$ and
exactly in this parameter for $\omega\gg m$, $Z|e|$ is the nucleus
charge, $\alpha =\, e^2/4\pi\, =1/137$ is the fine-structure
constant, $e$ and $m$ are the electron charge and mass,
$\hbar=c=1$. It turned out that higher orders of perturbation
theory with respect to $Z\alpha$ (Coulomb corrections) play an
important role and drastically modify the cross section of
Delbr\"uck scattering on heavy atoms at high photon energy. The
experience gained at the investigation of Delbr\"uck scattering
was extremely useful for the study of photon splitting. In
particular, the importance of the Coulomb corrections in the
latter was recognized.

In the process of photon splitting, the initial photon turns in
the electric field of an atom into two photons sharing its energy
$\omega_1$. For a long time, the amplitude of this process was
known only in the Born approximation \cite{Sh,CTP}. For
$\omega_1\ll m$, the asymptotic form of the Born amplitude was
derived earlier in \cite{Bol54,Bukh63}. First successful estimate
of the total cross section of high-energy ($\omega_1\gg m$) photon
splitting was made in \cite{Will35}. More accurate results were
obtained  in \cite{BKKF} within the same approach for the total as
well as for the partly integrated cross sections. The detailed
numerical investigation based on the results of \cite{Sh,CTP} was
performed in \cite{JMO80,St83}.

Recently, an essential progress in understanding of photon
splitting phenomenon was achieved due to the use of the
quasiclassical approach. In papers \cite{LMS1,LMS2,LMS3} various
differential cross sections of high-energy photon splitting have
been calculated exactly in the parameter $Z\alpha$. Similar to the
case of Delbr\"uck scattering, the exact cross section turns out
to be noticeably smaller than that obtained in the Born
approximation. So, the detailed theoretical and experimental
investigation of photon splitting provides a new sensitive test of
QED when the effect of higher-order terms of the perturbation
theory with respect to the external field is very important.

The observation of photon splitting is extremely hard problem due
to severe background conditions. The significance of various
competing processes depends on photon energy. For instance, at
$\omega_1\leq m$ the dominant background process is double Compton
scattering on atomic electrons. The search for photon splitting in
this energy region was performed in two experiments
\cite{Adler66,Rob66}.  As pointed out in \cite{Adler66,Rob66}, the
results of both experiments do not agree with the theoretical
estimates. Nevertheless, the authors of \cite{Adler66} argue that
their results strongly indicate the existence of photon splitting.
The energy region $\omega_1\gg m$ is more favorable for the
observation of the phenomenon, though the background is still
rather severe. Using the advantage of the intense source of tagged
photons, the first successful observation of photon splitting in
the energy region $120$~MeV$\leq\omega_1\leq 450$~MeV has been
performed recently, and the preliminary results are published in
\cite{Akh97}, \cite{ph98}. Here we present the  new data analysis
for this experiment \cite{ph2001}. The results obtained
 confirm the existence of photon
splitting phenomenon. They also  make possible  the quantitative
comparison with the theoretical predictions. Moreover, the
attained experimental accuracy allows one to distinguish between
the theoretical predictions obtained with or without accounting
for the Coulomb corrections. It turns out that the Coulomb
corrections essentially improve the agreement between the theory
and the experiment.

\section{General discussion}

Let a photon with 4-momentum $k_1$ ($k_1^0=\omega_1=|\bm{k}_1|$)
and polarization vector $\bm{e}_1$ turns in the electric field of
an atom into two photons with 4-momenta $k_{2,3}$
($k_{2,3}^0=\omega_{2,3}=|\bm{k}_{2,3}|$) and polarization vectors
$\bm{e}_{2,3}$. We assume that the momentum transferred to a
nucleus is small as compared to the mass of the nucleus. This
condition allows us to neglect the recoil effects and consider an
atom as a source of a time-independent electric field. Then, the
final photons share the energy of the initial quantum:
$\omega_1=\omega_2+\omega_3$. Later we will see that for arbitrary
photon energy the main contribution to the total cross section
comes from recoil momenta smaller than several tenth of electron
mass. Therefore, the approximation of an external field is valid
everywhere except the kinematic region where the differential
cross section of the process is negligibly small.

The most convenient way to take into consideration an external
electromagnetic field in quantum electrodynamics is to use the
Furry representation. In this approach the amplitude of a process
is described by a set of Feynman diagrams where the electron lines
correspond to the Green functions of the Dirac equation in the
field. As a result, we obtain the amplitude in the form of series
in the parameter $\alpha$, where the coefficients are exact in the
external field strength.

\begin{figure}[h]
\centering
 \includegraphics[height=100pt,keepaspectratio=true]{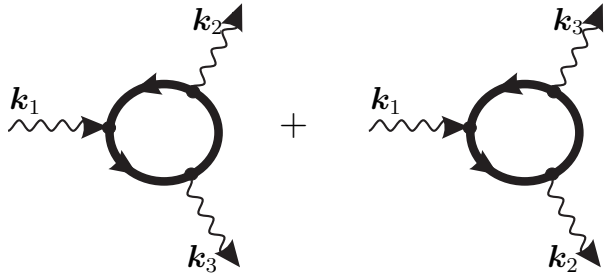}
 \begin{picture}(40,0)(0,0)
 \put(-25,0){$\bm{k}_3$}\put(-92,60){$\bm{k}_1$}
 \put(-23,90){$\bm{k}_2$}
 \put(10,50){\large$+$}
 \end{picture}
 \includegraphics[height=100pt,keepaspectratio=true]{3gamma.eps}
 \begin{picture}(0,0)(0,0)
 \put(-25,0){$\bm{k}_2$}\put(-92,60){$\bm{k}_1$}
 \put(-23,90){$\bm{k}_3$}
 \end{picture}
 \caption{Feynman diagrams for the photon splitting amplitude in the Furry
 representation. Thick lines denote the electron propagator in an external
 field}
 \label{ps1}
\end{figure}

Here we consider the amplitude $M$ of photon splitting only in the
lowest order in $\alpha$ which is given by the Feynman diagrams
shown in Fig. \ref{ps1}.

The corresponding analytical expression for $M$ reads
\begin{eqnarray}
\label{General:Amplitude} &&M={i}{e^3}\int\frac{d\varepsilon}{2\pi}\int
d\bm{r}_1\,d\bm{r}_2\,d\bm{r}_3
\exp[i(\bm{k}_1\cdot\bm{r}_1-\bm{k}_2\cdot\bm{r}_2-\bm{k}_3\cdot\bm{r}_3)]
 \\
&&\times\mbox{Tr}[\hat e_1
G(\bm{r}_1,\bm{r}_2|\varepsilon-\omega_2) \hat
e_2^*G(\bm{r}_2,\bm{r}_3|\varepsilon)\hat
e_3^*G(\bm{r}_3,\bm{r}_1|\varepsilon+\omega_3)]
+(k_2^{\mu}\leftrightarrow k_3^{\mu}, \bm{e}_2\leftrightarrow
\bm{e}_3)\, .\nonumber
\end{eqnarray}
Here $\hat e = e^{\mu}\gamma_{\mu}=-\bm{e}{\bm{\gamma}}$,
$\gamma^{\mu}$ being the Dirac matrices,
$G(\bm{r}_1,\bm{r}_2|\varepsilon)$ is the Green function of the
Dirac equation in the external electric field:
\begin{equation}
G(\bm{r}_1,\bm{r}_2|\varepsilon)=\langle \bm{r}_1|
[\hat{\mathcal{P}}-m+i0]^{-1}|\bm{r}_2\rangle\, , \quad
\hat{\mathcal{P}}=\gamma^0 (\varepsilon-U(\bm{r})) - \bm{\gamma}\cdot\bm{p}\,,
\end{equation}
$\bm{p}=-i\bm{\nabla}$, $U(\bm{r})$ is the potential energy of an
electron. Note that, according to the Furry theorem, the amplitude
(\ref{General:Amplitude}) is an odd function of the external field
strength (of the parameter $Z\alpha$ for a Coulomb field).

The differential cross section of photon splitting has the form
\begin{equation}
\label{General:CrossSection} d\sigma=\frac1{2^8\pi^5}
|M|^2\,\omega_1^2\, x(1-x)\, dx\,d\Omega_2\,d\Omega_3\, ,
\end{equation}
where $x=\omega_2/\omega_1$ (so that $\omega_3=\omega_1(1-x)$),
and $\Omega_{2,3}$ are solid angles of $\bm{k}_{2,3}$. It is
convenient to perform the calculations in terms of the helicity
amplitudes
$M_{\lambda_1\lambda_2\lambda_3}(\bm{k}_1,\bm{k}_2,\bm{k}_3)$,
with $\lambda_i=\pm 1$. In this case the polarization vectors
satisfy the relations $\bm{e}_{\lambda}\cdot \bm{k}=0$ and
$\bm{e}_{\lambda}\times\bm{k}=i\lambda\,\omega\bm{e}_{\lambda}$.
In fact, it is sufficient to calculate three amplitudes, e.g.,
$M_{+++}(\bm{k}_1,\bm{k}_2,\bm{k}_3)$,
$M_{++-}(\bm{k}_1,\bm{k}_2,\bm{k}_3)$, and
$M_{+--}(\bm{k}_1,\bm{k}_2,\bm{k}_3)$. Due to parity conservation
and identity of photons, other amplitudes can be obtained by the
substitutions (see, e.g. \cite{Sh}). The cross section
$d\sigma_{\lambda_1\lambda_2\lambda_3}$ for circularly polarized
photons is given by (\ref{General:CrossSection}) with $M$ being
$M_{\lambda_1\lambda_2\lambda_3}$. For unpolarized initial photon
the cross section summed up over the polarizations of final
photons is given by (\ref{General:CrossSection}) substituting for
$|M|^2$
\begin{equation}
\label{General:unpolarized} |M|^2\to \frac12\sum_{\lambda_i}
|M_{\lambda_1\lambda_2\lambda_3}|^2\,.
\end{equation}


Below we use the system of coordinates with $z$-axis directed
along $\bm{k}_1$ so that $a_z=\bm{a}\cdot\bm{k}_1/\omega_1$ and
$\bm{a}_\perp=\bm{a}-a_z\bm{k}_1/\omega_1$ for an arbitrary vector
$\bm{a}$. Then the amplitudes depend on $x=\omega_2/\omega_1$,
polar angles $\theta_{2,3}$ and azimuth angles $\phi_{2,3}$ of the
vectors $\bm{k}_{2,3}$. For spherically symmetric potential the
quantity $|M_{\lambda_1\lambda_2\lambda_3}|^2$ depends on azimuth
angles only via $\phi=\phi_2-\phi_3$. In this case, due to the
parity conservation, we have
\begin{equation}\label{General:Phi}
|M_{\lambda_1\lambda_2\lambda_3}(\phi)|^2=|M_{\Lambda_1\Lambda_2\Lambda_3}(-\phi)|^2\,,
\end{equation}
where $\Lambda_i$ denote the helicity opposite to $\lambda_i$.
Generally speaking, $|M_{\lambda_1\lambda_2\lambda_3}(\phi)|^2$ is
not an even function of $\phi$.

Four theoretical approaches, having different ranges of
applicability have been used for calculation of the cross section
of photon splitting in an atomic field. Three of them use the
first-order perturbation theory in the field strength, thereby
implying $Z\alpha\ll 1$. They are
\begin{itemize}
\item
The approach exploiting the Heisenberg-Euler effective Lagrangian,
which can be used only for $\omega_1\ll m$.
\item
The Weizs\"acker-Williams approximation, valid for $\omega_1\gg
m$, and providing the logarithmic accuracy for the cross section
of the process.
\item
The calculation of the amplitude exactly with respect to all
kinematic parameters of the problem. This approach will be
referred below as the Born approximation.
\end{itemize}

The fourth method is based on the quasiclassical approximation
valid for $\omega_1\gg m$. In contrast to the approaches itemized,
it gives the cross section exact in the parameter $Z\alpha$, and
provides a power accuracy in the small parameter $m/\omega_1$.

All these approaches will be reviewed below.


\section{Low-energy photon splitting}

The Heisenber-Euler effective Lagrangian (HEL), derived in
\cite{Hei35}, describes self-action of an electromagnetic field
due to the vacuum polarization when the momenta of quanta are
small compared to the electron mass $m$. This Lagrangian depends
on two invariants of the electromagnetic field:
$${\mathcal F}=\frac12(\bm{H}^2-\bm{E}^2)\quad ,\quad {\mathcal G}=\bm{E}\cdot\bm{H} \,.$$
The lowest order term of the expansion of the HEL with respect to
these invariants reads
\begin{equation}\label{HEL}
{\mathcal L} = \frac{e^4}{360\pi^2 m^4}\left[
(\bm{E}^2-\bm{H}^2)^2+7 (\bm{E}\cdot\bm{H})^2 \right]
\end{equation}
To calculate the photon splitting amplitude, it is necessary to
present the electromagnetic field in (\ref{HEL}) as a sum of a
quantized field and the classical electric field of an atom and
take the corresponding matrix element. The result obtained
accounts for a single Coulomb exchange and holds for $\omega_1\ll
m$ (and, hence, $\omega_{2,3}\ll m $) because the momentum
transfer $\bm{\Delta}=\bm{k}_2+\bm{k}_3-\bm{k}_1$ in this exchange
is also small as compared to $m$. We emphasize that it is
impossible to calculate the higher order corrections in $Z\alpha$
(Coulomb corrections) to the photon splitting amplitude using the
HEL. This is due to the fact that at the multiple Coulomb exchange
the typical momenta of individual Coulomb quanta are of the order
of $m$ though the sum of these momenta $\bm{\Delta}$ is small.

The calculation of the low-energy photon splitting amplitude with
the use of the HEL was performed in \cite{Bol54,Bukh63}.
Unfortunately, these papers contain some misprints in the
calculations of total cross section as was pointed out in
\cite{JMO80}, where the photon splitting was investigated
numerically using the amplitudes obtained in \cite{Sh} in the Born
approximation.

It follows from (\ref{HEL}) that the amplitude of the process in
an unscreened Coulomb field has the form
\begin{eqnarray}
\label{LowEnergy:Amplitude}
M_{\lambda_1\lambda_2\lambda_3}=\frac{Ze^5
\omega_1\omega_2\omega_3}{45\pi^2 m^4 \Delta^2}&& \biggl\{
(\bm{\Delta}\cdot\bm{e}_1)(\bm{e}_2^*\cdot\bm{e}_3^*)\left[1+\lambda_2
\lambda_3+\frac74\lambda_1 (\lambda_2+\lambda_3) \right]
\\
&&+
(\bm{\Delta}\cdot\bm{e}_2^*)(\bm{e}_1\cdot\bm{e}_3^*)\left[1-\lambda_1
\lambda_3+\frac74\lambda_2 (\lambda_1-\lambda_3) \right]\nonumber
\\&&+
(\bm{\Delta}\cdot\bm{e}_3^*)(\bm{e}_1\cdot\bm{e}_2^*)\left[1-\lambda_1
\lambda_2+\frac74\lambda_3 (\lambda_1-\lambda_2) \right]
 \biggr\} \nonumber\, .
\end{eqnarray}
In the case of the electric field of an atom this amplitude should
be multiplied by an atomic form factor $[1-F(\Delta^2)]$ which
accounts for screening. This form factor, being the Fourier
transform of a charge density (in units of $Z|e|$), vanishes at
$\Delta=0$ and tends to unity for $\Delta\to \infty$. As a
function of $\Delta$ it has a typical scale of $r_c^{-1}=m\alpha
z^{1/3}$. These features are illustrated by a simple
representation for the form factor suggested  in \cite{Mol47}
\begin{equation}\label{LowEnergy:FF}
1-F(\Delta^2)=\Delta^2\sum_{i=1}^{3}\,\frac{\alpha_{i}}{\Delta^2+\beta_{i}^2}
\, ,
\end{equation}
where \begin{eqnarray} \label{coef}
\alpha_{1}&=&0.1, \quad
\alpha_{2}=0.55 ,\quad \alpha_{3}=0.35 , \quad \beta_{i}=\beta_0
b_i ,
\\
\nonumber b_{1}&=&6 , \quad b_{2}=1.2 , \quad b_{3}=0.3 , \quad
\beta_0= mZ^{1/3}/121.
\end{eqnarray}
For $\omega r_c\ll 1$, the effect of screening leads to the strong
suppression of the cross section of photon splitting as compared
to the case of unscreened Coulomb field.

The amplitudes (\ref{LowEnergy:Amplitude}) obey the relation
$|M_{\lambda_1\lambda_2\lambda_3}|^2=|M_{\Lambda_1\Lambda_2\Lambda_3}|^2$
($\Lambda$ denotes the helicity opposite to $\lambda$). Then, as
follows from (\ref{General:Phi}), the quantity
$|M_{\lambda_1\lambda_2\lambda_3}|^2$ is an even function of
$\phi$.

\begin{figure}[h]
\centering
\includegraphics[height=200pt,keepaspectratio=true]{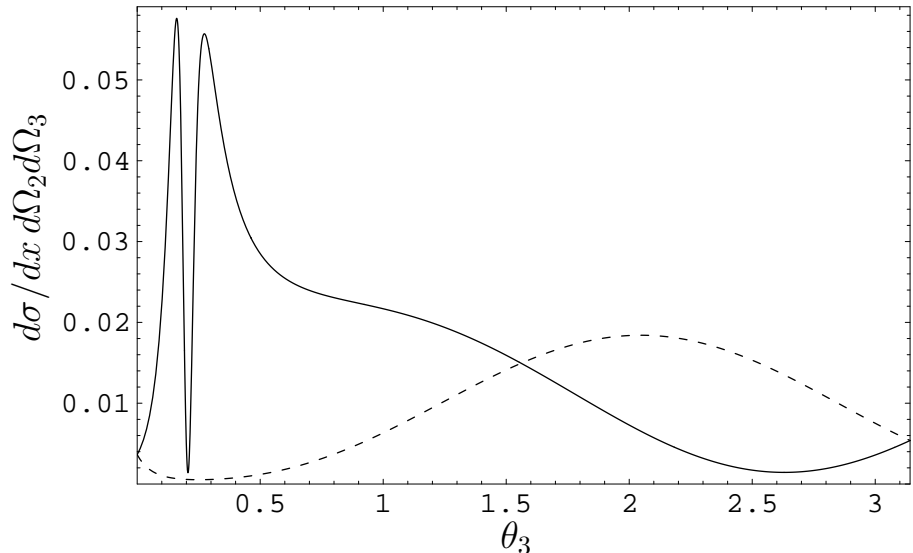}
\begin{picture}(0,-20)(0,0)
 \put(-160,-8){\large $\theta_3$}
 \put(-345,70){\large\rotatebox{90}{$d\sigma/dx\,d\Omega_2d\Omega_3$}}
 \end{picture}
\caption{The cross section $d\sigma/dx\,d\Omega_2d\Omega_3$ in
units of $10^{-4} Z^2\alpha^5{\omega_1}^6/{m^8}=3.09\cdot
Z^2(\omega_1/m)^6$ pb as a function of $\theta_3$ for $x=0.4$,
$\theta_2=\pi/10$, and $\phi=0,\pi$ (dashed and solid curves,
respectively). }
 \label{LowEnergy:SigDif(theta)}
\end{figure}

The angular distribution of the final photons is rather
complicated. This is illustrated in Fig.
\ref{LowEnergy:SigDif(theta)}, where the cross section
$d\sigma/dx\,d\Omega_2d\Omega_3$ is shown as a function of
$\theta_3$ at $\theta_2=\pi/10$ and $\phi=0,\pi$.

\begin{figure}[h]
\centering\includegraphics[height=180pt,keepaspectratio=true]{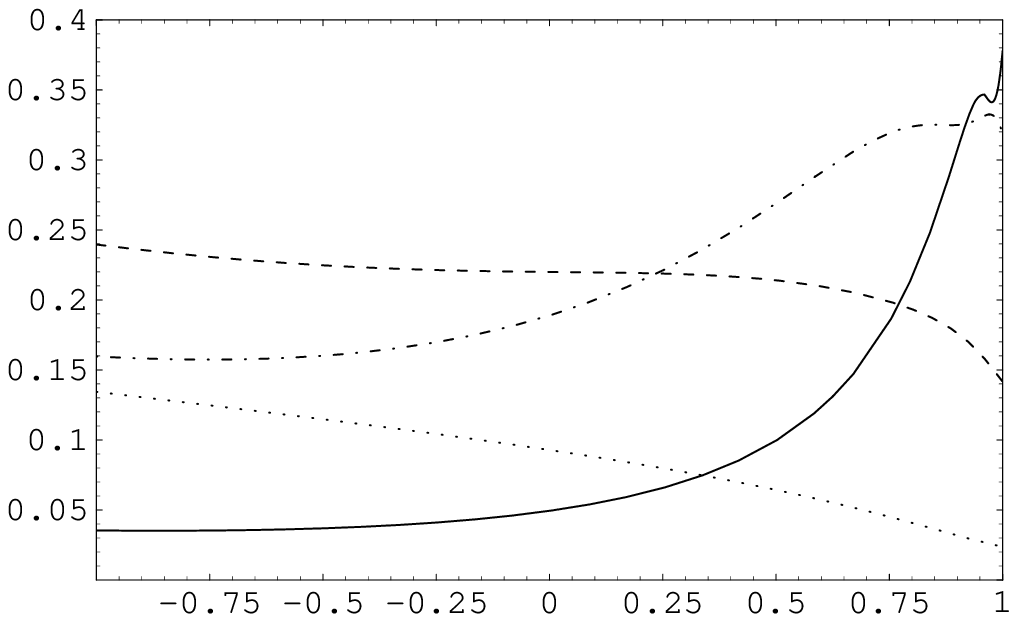}
\begin{picture}(0,0)(0,0)
 \put(-145,-8){\large $\cos \theta_2$}
 \put(-308,85){\large\rotatebox{90}{$d\sigma/dx\,d\Omega_2$}}
 \end{picture}
\caption{The cross section $d\sigma/dx\, d\Omega_2$ in the same
units as in Fig. \ref{LowEnergy:SigDif(theta)} versus $\cos
\theta_2$ for $x=0.2,0.4,0.6,0.8$ (dotted, dashed, dash-dotted,
and solid curves, respectively). }
 \label{LowEnergy:Sig(theta)}
\end{figure}

\begin{figure}[h]
\centering\includegraphics[height=200pt,keepaspectratio=true]{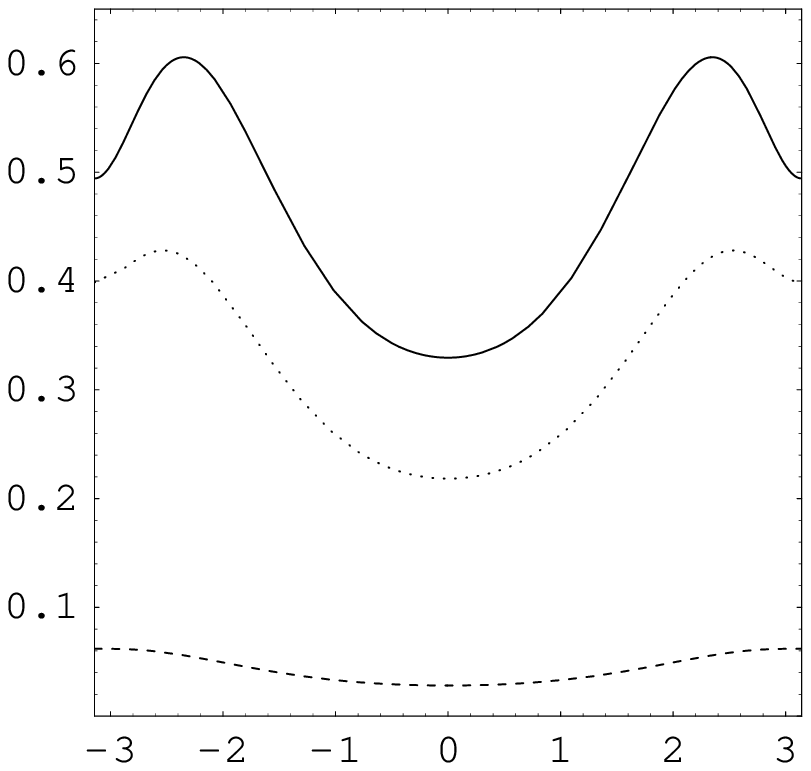}
\begin{picture}(0,0)(0,0)
 \put(-95,-8){\large $\phi$}
 \put(-218,85){\large\rotatebox{90}{$d\sigma/dx\,d\phi$}}
 \end{picture}
\caption{The cross section $d\sigma/dx\,d\phi$ in the same units
as in Fig. \ref{LowEnergy:SigDif(theta)} versus $\phi$ for
$x=0.1,0.3,0.5$ (dashed, dotted, and solid curves, respectively).
}
 \label{LowEnergy:Sig(phi)}
\end{figure}

\begin{figure}[h]
\centering\includegraphics[height=180pt,keepaspectratio=true]{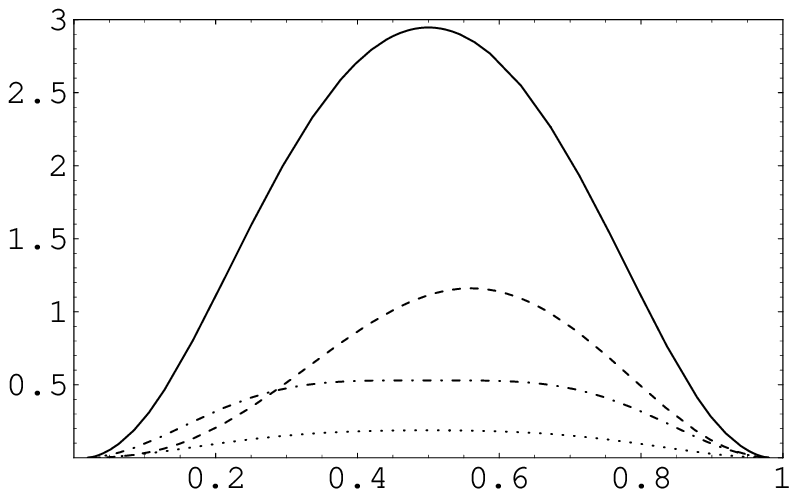}
\begin{picture}(0,0)(0,0)
 \put(-145,-4){\large $x$}\put(-308,85){\large\rotatebox{90}{$d\sigma/dx$}}
 \end{picture}
\caption{The cross section $d\sigma/dx$ in the same units as in
Fig. \ref{LowEnergy:SigDif(theta)}, for unpolarized photons (solid
curve) and for special cases of helicities:
$\lambda_1\lambda_2\lambda_3=+--$ (dotted curve),$++-$ (dashed
curve), and $+--$ (dash-dotted curve). }
 \label{LowEnergy:Sig(x)}
\end{figure}

For $\phi=\pi$ there is a double peak with a narrow notch. Such a
structure occurs in the range of the momentum transfer
$\Delta\ll\omega_1$ where $\theta_{2,3}\ll 1$. Then the bottom of
the notch is at $\phi=\pi$, $\theta_3=\theta_2 x/(1-x)$, when
$\Delta_\perp=0$. Note that for $\phi=0$ the differential cross
section has a wide deep around $\theta_3=\theta_2$. In fact, such
a suppression of the cross section, when the vectors $\bm{k}_2$
and $\bm{k}_3$ are almost parallel, persists for any $\bm{k}_2$.
The cross section $d\sigma/dx\, d\Omega_2$ differential with
respect to the momentum of one photon is plotted in Fig.
\ref{LowEnergy:Sig(theta)} for different values of $x$ as a
function of $\cos \theta_2$. It is seen that the photons with
$1-x\ll 1$ ($x\ll 1$) are emitted preferably along $\bm{k}_1$
($-\bm{k}_1$). It is also interesting to consider the dependence
of the cross section on the azimuth angle $\phi=\phi_2-\phi_3$,
shown in Fig. \ref{LowEnergy:Sig(phi)}. This cross section is
symmetric with respect to the replacement $x\to 1-x$. One can see
that for all $x$ the $\phi$-distribution has a minimum at
$\phi=0$. The cross section being the function of $|\phi|$ has a
maximum that moves from $|\phi|=\pi$ to $|\phi|=2.35$ when $x$
increases from $0$ to $0.5$. The cross sections
$d\sigma_{\lambda_1\lambda_2\lambda_3}/dx$ for different
helicities as well as $d\sigma/dx$ for unpolarized photons  are
shown in Fig. \ref{LowEnergy:Sig(x)}. Note that $d\sigma_{++-}/dx$
is not symmetric with respect to substitution $x\to 1-x$. Instead,
after this substitution $d\sigma_{++-}/dx$ turns into
$d\sigma_{+-+}/dx$.

Numerical integration gives the following result for the total
cross section in the Coulomb field
\begin{equation} \label{LowEnergy:CrossSection}
\sigma=\frac12\int\limits_0^1\frac{d\sigma}{dx}dx=7.6\cdot 10^{-5}
\frac{Z^2 \alpha^5}{m^2} \left(\frac{\omega_1}{m}\right)^6\, ,
\end{equation}
which coincides with that obtained in \cite{JMO80}. The factor
$1/2$ in front of the integral accounts for the identity of
photons.

\section{Weizs\"acker-Williams approximation}

The main contribution to the total cross section of photon
splitting for $\omega_1\gg m$ comes from the region of small
angles $\theta_{2,3}\sim m/\omega_1$ and can be obtained within
the logarithmic accuracy using the Weizs\"acker-Williams (WW)
approximation. Using this approximation, rather accurate estimate
of the total cross section was first made in \cite{Will35} and
later in \cite{Bol54,Bukh63}. The partially integrated cross
section as well as the total one were calculated in this
approximation in \cite{BKKF}. In the WW approximation the
interaction with a Coulomb quantum is replaced by that with a real
photon having the 4-momentum $q=(q^0,-q^0\bm{k}_1/\omega_1)$. Then
the cross section of photon splitting in WW approximation is
expressed via the cross section of photon-photon scattering
$d\sigma_{\gamma\gamma}$ as follows
\begin{eqnarray}\label{WW:CrossSection}
&&d\sigma=\frac{2Z^2\alpha}{\pi}\frac{ds}s L(s)
d\sigma_{\gamma\gamma}\,,\quad
L(s)=\int\limits_{\Delta_{min}}^{\Delta_{eff}}\frac{d\Delta}{\Delta}
[1-F(\Delta^2)]^2\, ,\\
&&s=(k_1+q)^2=4\omega_1q^0\, ,\quad
\Delta_{min}=\frac{s}{\omega_1}\, ,\quad
\Delta_{eff}=\sqrt{s}\nonumber\, ,
\end{eqnarray}
where $[1-F(\Delta^2)]$ is the atomic form factor
(\ref{LowEnergy:FF}). The factor in front of
$d\sigma_{\gamma\gamma}(s)$ is the spectral distribution of
equivalent photons. The photon-photon scattering cross section has
the form
\begin{equation}\label{WW:CrossSection2gamma}
d\sigma_{\gamma\gamma}=|\tilde{M}|^2 \delta(k_1+q-k_2-k_3)
\frac{d\bm{k}_2d\bm{k}_3}{32\pi^2 s\, \omega_2\omega_3}\, ,
\end{equation}
where $\tilde{M}$ is the amplitude of photon-photon scattering
\cite{CTP}. To eliminate the four-dimensional $\delta$-function in
the cross section, it is necessary to integrate over the momentum
of one of the final photons, and over $q_0$ in
(\ref{WW:CrossSection}). We have
$\bm{k}_{2\perp}+\bm{k}_{3\perp}=0$ or $\phi=\pi$ and
$\omega_2\theta_2=\omega_3\theta_3$ for small polar angles. Then
the differential cross section $d\sigma/dx d\bm{k}_{2\perp}$ has
the form
\begin{equation}\label{WW:CS(theta)}
\frac{d\sigma}{dxd\bm{k}_{2\perp}}=\frac{Z^2\alpha{|\tilde{M}|^2}
L(s)}{8 \pi^3x(1-x) s^2 }\, ,\quad
s=\frac{{k}_{2\perp}^2}{x(1-x)}\, .
\end{equation}

The differential cross section for unpolarized photons is given by
(\ref{WW:CS(theta)}) substituting for $|\tilde{M}|^2$
\begin{eqnarray}\label{WW:Eq:Mbar}
|\tilde{M}|^2&\to&\frac12[|\tilde{M}_{++++}|^2+|\tilde{M}_{++--}|^2+|\tilde{M}_{+-+-}|^2
+|\tilde{M}_{+--+}|^2+4|\tilde{M}_{+++-}|^2]\, ,
\end{eqnarray}
where $\tilde{M}_{\lambda_1\lambda_q\lambda_2\lambda_3}$ are the
helicity amplitudes of photon-photon scattering which explicit
form is given in \ref{AppendixWW}.

Integrating (\ref{WW:CS(theta)}) over all variables we obtain the
total cross section of photon splitting which can be represented
in the form \cite{BKKF}
\begin{eqnarray}\label{WW:CStotal}
\sigma&=&\frac{2 Z^2\alpha}\pi\int\limits_0^\infty
\frac{ds}s\sigma_{\gamma\gamma}(s) L(s)\,.
\end{eqnarray}

Since the cross section $\sigma_{\gamma\gamma}(s)$ falls off for
$s\gg m^2$, the integral in (\ref{WW:CStotal}) converges at $s\sim
m^2$. Within the accuracy provided by the WW method we can take
the quantity $L(s=m^2)$  outside the integral. The magnitude of
$L(m^2)$ strongly depends on the ratio of
$\Delta_{min}^{-1}=\omega_1/m^2$ to the radius of screening
$r_c\sim [m\alpha Z^{1/3}]^{-1}$. In the region $\omega_1 \geq
m^2r_c$ we have $L(m^2)=\ln mr_c=\ln (Z^{-1/3}/\alpha)$ and
$\sigma$ turns out to be independent of $\omega_1$. For
$m\ll\omega_1 \leq m^2r_c$, when the screening may be neglected,
we have $L(m^2)=\ln (\omega_1/m)$. As a result, for the total
cross section we have
\begin{eqnarray}\label{WW:CStotal1}
\sigma&=&0.725\frac{Z^2\alpha^5}{m^2}\times\left\{
  \begin{array}{ll}
    \ln (\omega_1/m)& \quad\mbox{for $m\ll\omega_1 \leq m Z^{-1/3}/\alpha $} \\
    &\\
    \ln (Z^{-1/3}/\alpha)& \quad\mbox{for $\omega_1 \geq m Z^{-1/3}/\alpha $}
  \end{array}
\right.
\end{eqnarray}

Recall that this result has the logarithmic accuracy. The
coefficient $0.725$ in (\ref{WW:CStotal1}) is taken from Eq.~(4.2)
of \cite{JMO80}. Instead of this number, the approximate value
$0.753$ was obtained in \cite{BKKF}.

The detailed comparison of the results obtained within WW
approximation, the Born approximation, and the quasiclassical
approximation will be performed below. Here we only note that the
accuracy of (\ref{WW:CStotal1}) at $\omega_1 \gg m Z^{-1/3}/\alpha
$ is about $12\%$ but for the differential  cross section
(\ref{WW:CS(theta)}) this accuracy may be essentially worse.

\section{Born approximation}

\begin{figure}[h]
 \centering\includegraphics[height=200pt,keepaspectratio=true]{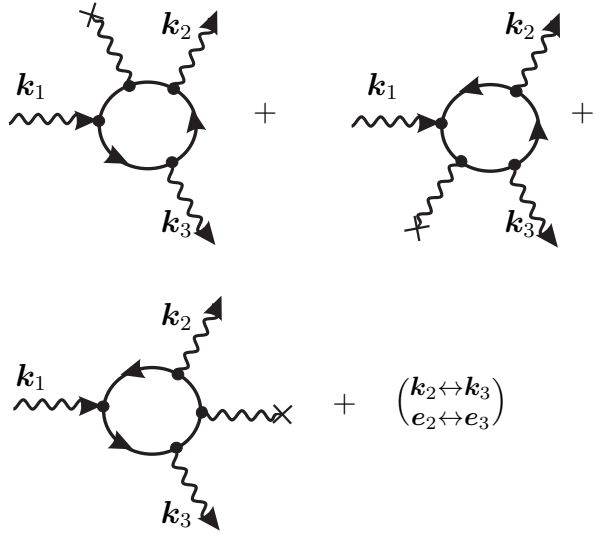}
 \begin{picture}(0,0)(0,0)
 \put(-210,165){$\bm{k}_1$}\put(-155,188){$\bm{k}_2$}\put(-155,113){$\bm{k}_3$}
 \put(-77,165){$\bm{k}_1$}\put(-25,188){$\bm{k}_2$}\put(-25,113){$\bm{k}_3$}
 \put(-210,55){$\bm{k}_1$}\put(-155,78){$\bm{k}_2$}\put(-155,3){$\bm{k}_3$}
\put(-120,155){$+$}\put(0,155){$+$} \put(-90,45){$+$\large
 $\quad{{ \bm{k}_2\leftrightarrow\bm{k}_3}
 \choose { \bm{e}_2\leftrightarrow\bm{e}_3}}$}
 \end{picture}
\caption{Feynman diagrams for the Born photon splitting amplitude.
Wavy lines with crosses denote Coulomb quanta.} \label{psB}
\end{figure}

In two previous sections we have considered photon splitting
within the first-order perturbation theory in $Z\alpha$ using some
additional approximations valid in restricted kinematic regions.
The analytical expression for the amplitudes in the Born
approximation valid for arbitrary momenta of photons was obtained
for the first time in \cite{Sh} and later confirmed in \cite{CTP}.
The Feynman diagrams corresponding to the amplitude in the Born
approximation are shown in Fig. \ref{psB}.

For unpolarized photons the result of \cite{Sh}, multiplied by the
atomic form factor, reads
\begin{equation}
\label{Born:Eq:Shima} {d\sigma\over dx\,d\Omega_2\,d\Omega_3}=Z^2\alpha^5
{[1-F(\Delta^2)]^2}
\frac{x(1-x)\omega_1^2}{2\pi^4(-\zeta\beta\gamma){\Delta^4}} \,\sum_{i=1}^4
(|x_i|^2+\eta |y_i|^2)\, ,
\end{equation}
where
\begin{eqnarray}
\zeta&=&-2k_2k_3/m^2\,,\ \beta=2k_1k_2/m^2\,,\
\gamma=2k_1k_3/m^2\,,\
\eta=4(\bm{k}_1\cdot[\bm{k}_2\times\bm{k}_3])^2/m^6\, .
\end{eqnarray}
The quantities $x_i$ and $y_i$ depend on $\omega_i$ and invariants
$\zeta$, $\beta$, $\gamma$. Their explicit form being very
cumbersome is presented in \ref{AppShima}.

\begin{figure}[h]
  \centering
    \includegraphics[height=180pt,keepaspectratio=true]{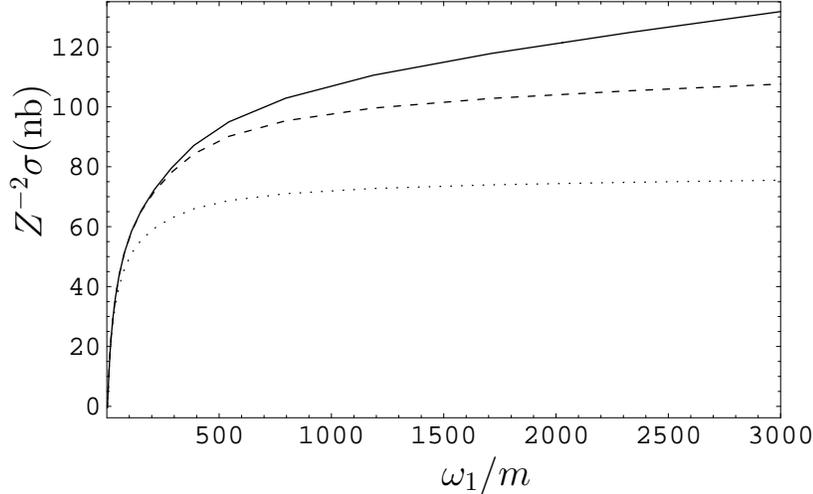}
\begin{picture}(0,0)(0,0)
 \put(-145,-8){\large $\omega_1/m$}
 \put(-308,85){\large\rotatebox{90}{$Z^{-2}\sigma \mathrm{(nb)}$}}
 \end{picture}
  \caption{The cross section $\sigma/Z^2\mathrm{(nb)}$  plotted as
a function of $\omega_1$ for a Coulomb potential (solid curve),
and the screened potential at $Z=1$ and $Z=92$ (dashed and dotted
curves, respectively).}
  \label{B:Fig:Tot}
\end{figure}

The numerical investigation of various differential cross sections
of photon splitting, based on the analytical result of \cite{Sh},
was performed in \cite{JMO80}. The results of \cite{JMO80} are
quoted in this section. Shown in Fig. \ref{B:Fig:Tot} is the
dependence of the total Born cross section on $\omega_1$ for
screened and unscreened Coulomb potentials. The values of this
cross section are listed in Table \ref{AppendixB:Table1} of
\ref{AppendixB} for various $Z$ and $\omega_1$.

 The numerical results for the total cross section at
$\omega_1\leq 0.1m$, obtained in \cite{JMO80}, coincide with
(\ref{LowEnergy:CrossSection}). The analytical fit to the cross
section in a Coulomb field, reproducing the cross section in the
range $10m\leq\omega_1\leq3000m$ within $0.1\%$, reads
\begin{equation}
\label{B:Eq:Fit} \sigma_{unscr}=Z^2 \left[22.372
\ln\frac{\omega_1}m -47.313
+82.536\frac{m}{\omega_1}-31.64\left(\frac{m}{\omega}\right)^2
\right]\, \mathrm{nb}\, .
\end{equation}

In contrast to the case of a  Coulomb field, when the total cross
section grows logarithmically with increasing $\omega_1$, the
photon splitting cross section in the case of a screened Coulomb
field is saturated for $\omega_1\gg m Z^{-1/3}/\alpha$. This trend
is clearly seen in Fig. \ref{B:Fig:Tot}. Note that the difference
of the cross sections  for screened and  Coulomb fields at
$\omega_1=m Z^{-1/3}/\alpha$ is small (a few percent).

The quasiclassical approach presented in the next Section allows
one to obtain the photon splitting cross section for $\omega\gg m$
with a power accuracy in the parameter $m/\omega_1$. Therefore,
the results obtained within this approximation reproduce for
$Z\alpha\ll 1$ two first terms of (\ref{B:Eq:Fit}). Thus, the
accuracy of the quasiclassical approach can be estimated by the
relative magnitude of two last terms in (\ref{B:Eq:Fit}). This
accuracy turns out to be better than $1\%$ at $\omega_1>70$MeV.

\begin{figure}[h]
  \centering
    \includegraphics[height=180pt,keepaspectratio=true]{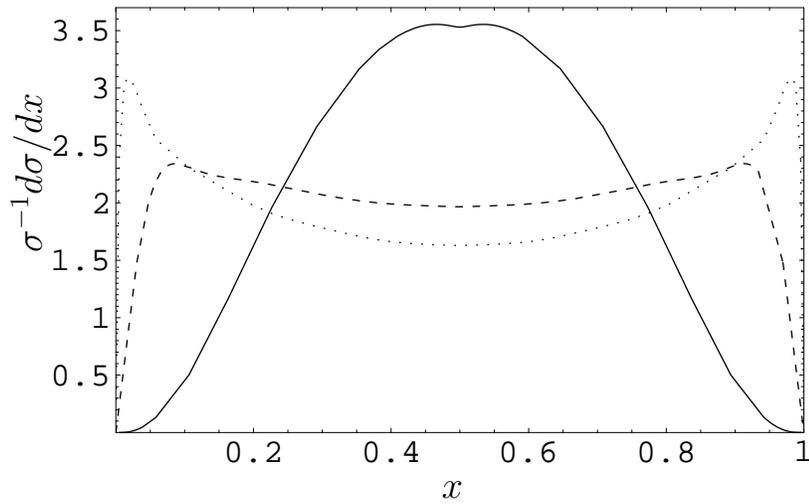}
\begin{picture}(0,0)(0,0)
 \put(-145,-8){\large $x$}
 \put(-308,85){\large\rotatebox{90}{$\sigma^{-1}d\sigma/dx$}}
 \end{picture}
  \caption{The spectrum $\sigma^{-1}d\sigma/dx$ is plotted for
the case of a Coulomb potential as a function of $x$ for
$\omega_1/m=1,10,100$ (solid, dashed, and dotted curves,
respectively).}
  \label{B:Fig:Spectr}
\end{figure}

In Fig. \ref{B:Fig:Spectr} the spectrum $\sigma^{-1}d\sigma/dx$
for a Coulomb field is plotted as a function of $x$ for different
values of $\omega_1$. One can see that the shape of the spectrum
changes significantly with increasing photon energy. The detailed
data obtained in \cite{JMO80} for spectrum in screened and
 Coulomb fields are presented in \ref{AppendixB} in
the Table \ref{AppendixB:Table2}.

In order to realize the accuracy of the Born approximation at
$Z\alpha\sim 1$ it is necessary to calculate the Coulomb
corrections. This issue is elucidated in the next Section.

\section{High-energy photon splitting}
\label{Chapter:HighEnergy}

Small scattering angles of particles provide the applicability of
the quasiclassical approach to the description of various
high-energy processes in an external field. This approach gives
the transparent physical picture of phenomena and essentially
simplifies calculations. Recall that the region of small angles
$\theta_{2,3}\ll 1$ makes the main contribution to the total cross
section of photon splitting for $\omega_1\gg m$. That is why we
confine ourselves on this kinematic region when considering the
process in this Section. Below we quote the results obtained in
the quasiclassical approximation in \cite{LMS1,LMS2,LMS3}. The
technique used in these papers was developed in \cite{MS83,LM95}
for the investigation of coherent photon scattering in a Coulomb
field.

We choose the origin at the center of a potential and direct the
$z$-axis along $\bm{k}_1$. Then characteristic values of
transverse components of $\bm{r}_i$ making the main contribution
to the integral in (\ref{General:Amplitude}) at given momentum
transfer $\bm{\Delta}=\bm{k}_2+\bm{k}_3-\bm{k}_1$ are of the order
of $\rho\sim 1/\Delta$. At small angles the characteristic angular
momentum is $l\sim \omega/\Delta \gg 1$, and the quasiclassical
approximation can be applied.

It is important that for $\omega\gg m$ the exact in $Z\alpha$
amplitude of photon splitting in the atomic field can be obtained
from that calculated for a Coulomb field. For this purpose, we
represent the amplitude as a sum of the lowest in $Z\alpha$ term
(Born amplitude) and the Coulomb corrections. As explained above,
the Born amplitude of the process in the atomic field is a product
of that in the Coulomb field and the atomic form factor. In the
perturbation theory the Coulomb corrections correspond to the
multiple interaction with quanta of an external field. The sum of
the momenta of these quanta equals $\bm{\Delta}$. It turns out
that even for $\Delta\ll m$ the characteristic magnitude of the
momentum of each quantum is not small as compared to $m$. As $m\gg
r_c^{-1}$ all form factors appearing in the calculation of the
Coulomb corrections should be set to unity. Thereby, the Coulomb
corrections are the same for the atomic and Coulomb fields.

\subsection{Quasiclassical picture of the process}

The calculation of the amplitude of photon splitting is
essentially simplified by using the Green function
$D(\bm{r}_1,\bm{r}_2|\varepsilon)$ of the ``squared'' Dirac
equation:
\begin{equation}
D(\bm{r}_1,\bm{r}_2|\varepsilon)=\langle
\bm{r}_1|[\hat{\mathcal{P}}^2-m^2+i0]^{-1}|\bm{r}_2\rangle\, .
\end{equation}
The Green function $G(\bm{r}_1,\bm{r}_2|\varepsilon)$ of the Dirac
equation can be obtained from $D(\bm{r}_1,\bm{r}_2|\varepsilon)$
by means of the relation
\begin{equation}
\label{HEnergy:Eq:GviaD}
 G(\bm{r}_1,\bm{r}_2|\varepsilon)=
 [\gamma^0 (\varepsilon-U(\bm{r}_1)) + i \bm{\gamma}\cdot\bm{\nabla}_1+m]
 D(\bm{r}_1,\bm{r}_2|\varepsilon)\,.
\end{equation}

As shown in \cite{LMS1,LMS2}, the initial expression for the
photon splitting amplitude (\ref{General:Amplitude}) can be
presented as a sum of two contributions, containing either three
or two Green functions $D(\bm{r}_1,\bm{r}_2|\varepsilon)$:
$M=M^{(3)}+M^{(2)}$. The term $M^{(3)}$ is given by
\begin{eqnarray}\label{HighEnergy:M3}
M^{(3)}&=&\frac{i}{2}{e^3}\int\frac{d\varepsilon}{2\pi}\int
d\bm{r}_1d\bm{r}_2d\bm{r}_3
\exp[i(\bm{k}_1\cdot\bm{r}_1-\bm{k}_2\cdot\bm{r}_2-\bm{k}_3\cdot\bm{r}_3)]
\nonumber\\
&& \times\mbox{Tr}\{[(-\hat e_1\hat k_1-2\bm{e}_1\cdot\bm{p})
D(\bm{r}_1,\bm{r}_2|\varepsilon-\omega_2)][(\hat e_2^*\hat
k_2-2\bm{e}_2^*\cdot\bm{p})D(\bm{r}_2,\bm{r}_3|\varepsilon)]
\nonumber \\
&& \times [(\hat e_3^*\hat
k_3-2\bm{e}_3^*\cdot\bm{p})D(\bm{r}_3,\bm{r}_1|\varepsilon+\omega_3)]\}
+(k_2^{\mu}\leftrightarrow k_3^{\mu}, \bm{e}_2\leftrightarrow\bm{e}_3) .
\end{eqnarray}

Here the operator $\bm{p}=-i\mbox{\boldmath
${\nabla}$\unboldmath}$ differentiates the Green functions $D$
with respect to their first argument. The term $M^{(2)}$ reads
\begin{eqnarray}\label{HighEnergy:M2}
M^{(2)}&=&i{e^3}\int\frac{d\varepsilon}{2\pi}\int d\bm{r}_1d\bm{r}_2\,
\mbox{Tr}\Bigl\{\exp[i(\bm{k}_1\cdot\bm{r}_1-\bm{k}_2\cdot\bm{r}_2
-\bm{k}_3\cdot\bm{r}_2)]
\nonumber
\\&&
\times\bm{e}_2^*\cdot\bm{e}_3^*[(-\hat e_1\hat
k_1-2\bm{e}_1\cdot\bm{p})D(\bm{r}_1,\bm{r}_2|\varepsilon-\omega_1)]
D(\bm{r}_2,\bm{r}_1|\varepsilon)\nonumber
\\&&
+\Bigl[\exp[i(\bm{k}_1\cdot\bm{r}_1-\bm{k}_2\cdot\bm{r}_2-\bm{k}_3\cdot\bm{r}_1)]
\bm{e}_1\cdot\bm{e}_3^*D(\bm{r}_1,\bm{r}_2|\varepsilon-\omega_2) \nonumber
\\&&
\times[(\hat e_2^*\hat k_2-2\bm{e}_2^*\cdot\bm{p})
D(\bm{r}_2,\bm{r}_1|\varepsilon)]
 + (k_2^{\mu}\leftrightarrow k_3^{\mu}
, \bm{e}_2\leftrightarrow\bm{e}_3)\Bigr]\Bigr\} .
\end{eqnarray}

Below the Green function $D(\bm{r}_1,\bm{r}_2\,|\varepsilon)$ will
be called the ``electron Green function'' for $\varepsilon > 0$
and the ``positron Green function'' for $\varepsilon < 0$.
According to \cite{MS83,LM95}, for high energies the  main
contribution to the integrals in (\ref{HighEnergy:M3}) and
(\ref{HighEnergy:M2}) comes from the region of integration over
the variables $z_i$ ($z$-coordinates of $\bm{r}_i$) in which $z'\,
<\, z$ for the electron Green function
$D(\bm{r},\bm{r}\,'|\varepsilon)$ and $z'\,>\, z$ for the positron
Green function. In terms of non-covariant perturbation theory,
this corresponds to the contribution of intermediate states (see
Fig. \ref{ps1}) for which the difference between their energy
$E_n$ and the energy  of the initial state $E_0=\omega_1$ is small
as compared to $E_0$.

\begin{figure}[h]
  \centering
  \includegraphics[height=150pt,keepaspectratio=true]{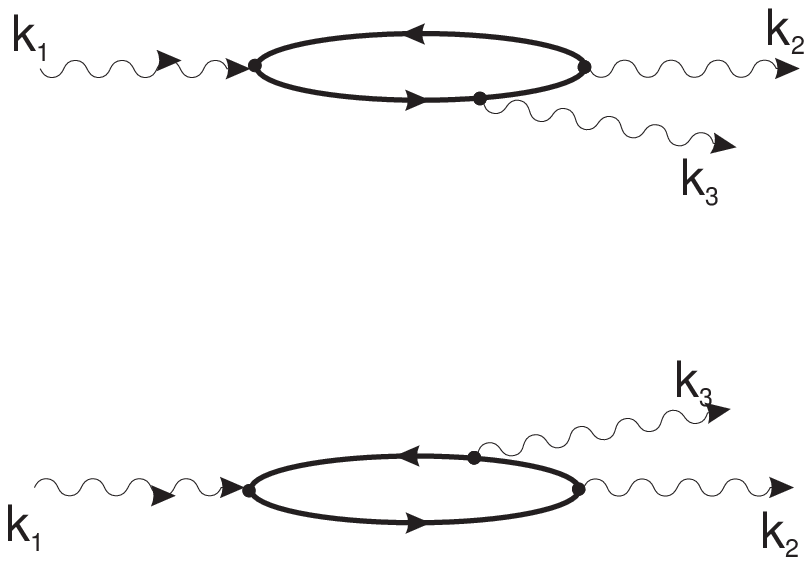}
  \hspace{40pt}
  \includegraphics[height=150pt,keepaspectratio=true]{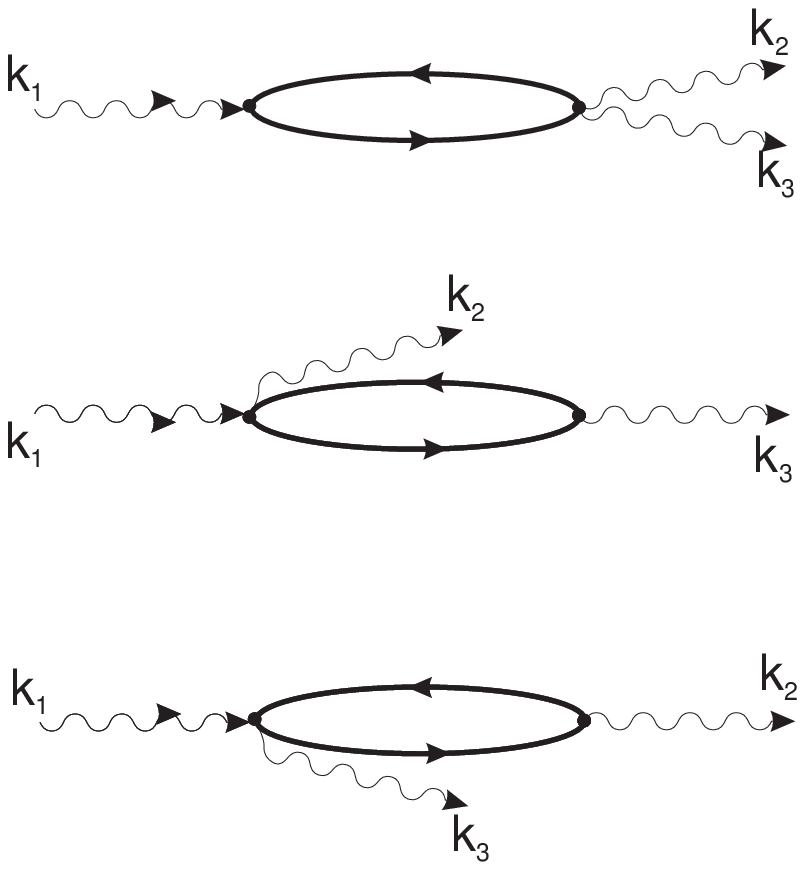}
 \begin{picture}(0,0)(0,0)
 \put(-392,22)
 { $+(k_2\leftrightarrow k_3, \bm{e}_2\leftrightarrow\bm{e}_3)$}
 \end{picture}
  \caption{Diagrams corresponding to the terms
  $M^{(3)}$ (left column) and
  $M^{(2)}$ (right column).}
  \label{HighEnergy:Fig:Diagrams}
\end{figure}

The lifetime of the intermediate state in photon splitting can be
estimated with the help of the uncertainty relation as $\tau\sim
\omega_1/(m^2+k_\perp^2)$, where
$k_\perp=\mbox{max}(|\bm{k}_{2\perp}|,\,|\bm{k}_{3\perp}|)$. The
characteristic transverse distance  between the virtual particles
(transverse size of the loop) can be estimated as
$(m^2+k_\perp^2)^{-1/2}$, which for $\omega_1\gg
\mbox{max}(m,k_\perp)$ is much smaller than the length of the loop
$\tau$. The quantity $\rho\sim1/\Delta$ introduced above can be
interpreted as a typical transverse distance between the loop and
the origin. In the small-angle approximation ($\theta_{2,3}\ll 1$)
we have for $\Delta^2$
\begin{equation}\label{HighEnergy:Eq:Delta}
{\Delta}^2=\bm{\Delta}_\perp^2+\Delta_z^2=\,
(\bm{k}_{2\perp}+\bm{k}_{3\perp}\,)^2+\frac{1}{4}
\left(\frac{\bm{k}_{2\perp}\,^2}{\omega_2}+
\frac{\bm{k}_{3\perp}\,^2}{\omega_3}\right)^2\, .
\end{equation}
It is seen from (\ref{HighEnergy:Eq:Delta}) that $\Delta_\perp\gg
\Delta_z$ everywhere except a narrow region where $\Delta_\perp\ll
k_\perp$. As pointed out in \cite{LMS2}, the amplitude at
$\Delta_\perp\sim \Delta_z$ can be obtained from that derived for
$\Delta_\perp\gg \Delta_z$ by a simple substitution that will be
discussed below. The condition $\Delta_\perp\gg \Delta_z$ implies
that $\rho/\tau\ll 1$ and, therefore, the angles between
$\bm{k}_i$ and $\bm{r}_i$ in (\ref{HighEnergy:M3}) and
(\ref{HighEnergy:M2}) are either small or close to $\pi$. The
additional restrictions on the region, making the main
contribution to the amplitude for $\Delta_\perp\gg \Delta_z$,
follow from the explicit form of the quasiclassical Green
function. For the term $M^{(3)}$ (\ref{HighEnergy:M3}) they are
\[
z_1<\, z_2 , z_3 \quad ,\quad z_1<0 \quad ,\quad \mbox{max}(z_2,\,
z_3)>\,0 \, .
\]
Similarly, the main contribution to the term $M^{(2)}$ (\ref{HighEnergy:M2}) is
given by the region $z_1 < 0$ and $z_2 > 0$. All these conditions allow one to
depict the main contribution to the amplitude $M^{(3)}$ and $M^{(2)}$ in the
form of diagrams, shown  in Fig.  \ref{HighEnergy:Fig:Diagrams}.

The explicit form of vertices is obvious from
(\ref{HighEnergy:M3}) and (\ref{HighEnergy:M2}). The electron
Green functions are marked with left-to-right arrows, and positron
ones with right-to-left arrows. The arrangement of the diagram
vertices is space ordered. With the use of these diagrams one can
easily determine the limits of integration over the energy and
coordinates. Diagrams in Fig. \ref{HighEnergy:Fig:Diagrams} have a
transparent interpretation. For example, the upper left diagram
corresponds to the following picture: the photon with momentum
$\bm{k}_1$ produces, at the point $\bm{r}_1$, a pair of virtual
particles which is transformed at the point $\bm{r}_2$ into a
photon with momentum $\bm{k}_2$. Between these two events the
electron emits a photon with the momentum $\bm{k}_3$ at the point
$\bm{r}_3$.

The quasiclassical Green function
$G(\bm{r}_1,\bm{r}_2|\varepsilon)$ of the Dirac equation in a
Coulomb potential was first derived in \cite{MS83} starting from
the exact Green function of the Dirac equation \cite{MS82}. For
the case of almost collinear vectors $ \bm{r}_1 $ and $ \bm{r}_2 $
the quasiclassical Green function in arbitrary spherically
symmetric decreasing potential was found in \cite{LM95}. Later in
\cite{LMS00} the quasiclassical Green function was obtained in an
arbitrary localized potential which generally possesses no
spherical symmetry.

As explained above for the calculation of the amplitude of photon
splitting in atomic field it is sufficient to know the
quasiclassical Green function $D(\bm{r}_1,\bm{r}_2|\varepsilon)$
of the ``squared'' Dirac equation in a Coulomb field for the case
of almost collinear vectors $ \bm{r}_1 $ and $ \bm{r}_2 $. The
corresponding expression derived in \cite{LMS1,LMS2} has the form
\begin{eqnarray}\label{HighEnergy:Eq:D1}
 D(\bm{r}_1,\bm{r}_2 |\,\varepsilon )&=& \,\frac{i\mbox{e}^{i\kappa
(r_1+r_2)}}{8\pi^2\kappa r_1r_2} \int d\bm{q}\,
\left[1+Z\alpha\frac{\bm{\alpha}\cdot\bm{q}}{q^2}\,
\right] \nonumber\\
&&\times \exp{\left [i\frac{q^2(r_1+r_2)}{2|\varepsilon|
r_1r_2}+i\bm{q}\cdot(\bm{\vartheta}_1+\bm{\vartheta}_2)\right]}
\left(\frac{4\kappa^2r_1r_2}{q^2}\right)^{iZ\alpha\eta}
\end{eqnarray}
for a small angle between vectors $\bm{r}_1$ and $-\bm{r}_2$. Here
$\bm{\alpha} = \gamma^0\bm{\gamma}$, $\kappa^2=\varepsilon^2-m^2$,
$\eta= \mathrm{sgn}(\varepsilon)$,
$\bm{\vartheta}_{1,2}=(\bm{r}_{1,2})_{\perp}/r_{1,2}$, and
$\bm{q}$ is a two-dimensional vector lying in $xy$ plane. The
direction of $z$ axis is arbitrary provided that
$\vartheta_{2,3}\ll 1$. Expression (\ref{HighEnergy:Eq:D1})
contains only elementary functions, and the angles
$\bm{\vartheta}_1$ and $\bm{\vartheta}_2$ appear only in the
factor $\exp[i\bm{q}\cdot(\bm{\vartheta}_1+\bm{\vartheta}_2)]$.
For a small angle between vectors $\bm{r}_1$ and $\bm{r}_2$ we
have for $D(\bm{r}_1,\bm{r}_2 |\,\varepsilon )$
\begin{equation}\label{HighEnergy:Eq:D2}
D(\bm{r}_1,\bm{r}_2 |\,\varepsilon )=
-\frac{\e^{i\kappa|\bm{r}_1-\bm{r}_2|}}{4\pi |\bm{r}_1-\bm{r}_2|}\,
\left(\frac{r_1}{r_2}\right)^{iZ\alpha\,\eta\,\mathrm{sgn}(r_1-r_2)}\, .
\end{equation}
One can see that the Green function (\ref{HighEnergy:Eq:D2})
differs from that at vanishing potential only by a phase factor.
It is easy to check that after the substitution of the expressions
(\ref{HighEnergy:Eq:D1}) and (\ref{HighEnergy:Eq:D2}) for the
Green functions to the splitting amplitudes (\ref{HighEnergy:M3})
and (\ref{HighEnergy:M2}) all phase factors of the form $r^{\pm
iZ\alpha}$ cancel.

\subsection{Amplitude of the process}

At the calculation of the amplitude it is convenient to eliminate
the $z$ component of the polarization vectors $\bm{e}_i$ using the
relation $\bm{e}_i\cdot\bm{k}_i=0$ which leads to
$e_{iz}=-\bm{e}_{i\perp}\cdot\bm{k}_{i\perp}/k_{iz}$.  After that,
in the small-angle approximation we can neglect the difference
between the polarization vector $\bm{e}_{2\perp}$
($\bm{e}_{3\perp}$) and that of photon, propagating along the $z$
axis and having the same helicity. As a result, we obtain the
amplitudes
$M_{\lambda_1\lambda_2\lambda_3}(\bm{k}_1,\bm{k}_2,\bm{k}_3)$
containing only the transverse polarization vectors $\bm{e}$ and
$\bm{e}^*$ for positive and negative helicities, respectively.
Though the Green function $D(\bm{r}_1,\bm{r}_2|\varepsilon)$
(\ref{HighEnergy:Eq:D1}), (\ref{HighEnergy:Eq:D2}) has a simple
form, the integration in (\ref{HighEnergy:M3}) and
(\ref{HighEnergy:M2}) was very sophisticated. The final
expressions, obtained in \cite{LMS3}, for the helicity amplitudes
exact in the parameter $Z\alpha$ have the form
\begin{eqnarray}
\label{HighEnergy:Eq:final} M_{+--} &=&N\!\!\int\!\! d{\bm x}\left(
{\bm{e}\cdot\bm{G}}\right)\!\! \int_0^{\omega _2}\!\! \frac{d\varepsilon
}A\kappa _2\Biggl[ - {\bm{e}\cdot\bm{a}} \left( \frac \varepsilon {
{\bm{e}}^*\cdot{\bm{\theta}}_3 }+\frac{\kappa _3}{
 {\bm{e}}^*\cdot{\bm{\theta}}_{23} }\right) \nonumber
\\
&& + m^2\omega _3\kappa _2\left( \frac{\kappa _3\,
{\bm{e}\cdot\bm{\theta}}_{23} }{\omega_1{\mathcal D}_1
{\bm{e}}^*\cdot{\bm{\theta}} _{23} }-\frac{\varepsilon \, {\bm
{e}\cdot\bm{\theta}}_3 }{\omega_2 {\mathcal D}_3 {\bm{e}}^*\cdot{\bm{\theta}}_3
}\right) \Biggr] +\left( \begin{array}{c} \omega_2\leftrightarrow \omega_3 \\
{\bm{\theta}}_2\leftrightarrow {\bm{\theta}}_3
\end{array}
\right) ,\nonumber\\
M_{+++} &=&N\!\!\int\!\! d{\bm x}\!\!\int_0^{\omega _2}\frac{d\varepsilon
}{2A}\left[ \frac{ {\bm{e}\cdot\bm{G}} \left( \kappa _2^2+\kappa _3^2\right) }{
{\bm{e}\cdot\bm{\theta}}_{23} }\left(  {\bm{e}}^*\cdot{\bm a} -\frac{ m^2\omega
_3\kappa _2 \,{\bm{e}}^*\cdot{\bm{\theta}}_{23} }{\omega _1 {\mathcal
D}_1}\right) \right.
\nonumber\\
&&  +\frac{\omega _1\omega _3\kappa _2G_1{\bm{e}}^*\cdot\left(
\bm{a}+\bm{\theta}_{23}\omega _2\kappa _3/\omega _1\right) }{{\mathcal D}_1}
+\frac{\omega _3\kappa _2 \,{\bm{e}}^*\cdot{\bm G} }{\omega _2{\mathcal D}_3}
\Bigl[2\varepsilon \left( \kappa _2^2+\kappa _3^2\right) \left( {\bm{e}}^*\cdot
{\bm{\theta}}_3\right) \left( {\bm{e}\cdot\bm{a}}\right)
\nonumber\\
&&\left.+ \varepsilon \omega _3A +m^2(2\varepsilon \kappa _2-\omega _1\omega
_2)\Bigr] -\frac{\omega _1\omega _3\kappa _2G_1\, {\bm{e}}^*\cdot{\bm c}
 }{{\mathcal D}_3}\right]
 +\left(\begin{array}{c}
\omega _2\leftrightarrow \omega _3 \\
{\bm{\theta}}_2\leftrightarrow {\bm{\theta}}_3
\end{array}  \right), \nonumber\\
M_{++-} &=&N\int d{\bm x\,}\Biggl\{ \int_0^{\omega _2}\frac{d\varepsilon }
{2A}\left[ \frac{\kappa _2\omega _3\, {\bm e}\cdot\bm{G} }{\omega _1 {\mathcal
D}_1}\Bigl[\kappa _3\left( \kappa _2-\varepsilon \right) A\right.
 -2\kappa_3\left( \kappa_2^2+\varepsilon ^2\right) \left( {\bm e}\cdot\bm{\theta}
 _{23}\right) \left(
{\bm{e}}^*\cdot{\bm a}\right)  \nonumber
\\
&&+m^2\left( \omega _1\omega _2-2\kappa _2\kappa _3\right) \Bigr]
 +\frac{\kappa _2\omega _2\omega _3G_1\,
{\bm e}\cdot\bm{b} }{{\mathcal D}_1} + \frac{\left( \kappa _2^2+\varepsilon
^2\right) {\bm{e}}^*\cdot {\bm G} }{ {\bm{e}}^*\cdot{\bm{\theta}}_3 }   \nonumber\\
&& \left.\times\left( {\bm e}\cdot\bm{a} +\frac{m^2\omega _3\kappa _2\, {\bm
e}\cdot\bm{\theta}_3 }{ \omega _2{\mathcal D}_3}\right) -\frac{\omega _2\omega
_3\kappa _2G_1{\bm{e}\cdot }\left( {\bm a}+\bm{\theta}_3\omega _1\varepsilon
/\omega _2\right) }{{\mathcal D}_3} \right]
\nonumber\\
&& +\int_{-\omega _3}^0d\varepsilon \frac{\omega _2\kappa _3}{2B}\left[ \frac{
{\bm e}\cdot\bm{G} }{\omega _1{\mathcal D}_1}\Bigl[-\left( \kappa
_2^2+\varepsilon \kappa _3\right) B+2\kappa _3\left( \kappa _2^2+\varepsilon
^2\right) \left( {\bm{e}}^*\cdot{\bm{\theta}}_{23}\right) \left( {\bm
e}\cdot\bm{b} \right) \right.
 \nonumber\\
&& +m^2\left( \omega _1\omega _2-2\kappa _2\kappa _3\right) \Bigr]
+\frac{\omega _2G_1\, {\bm e}\cdot\bm{b} }{{\mathcal D}_1}+\frac{\omega _2G_1\,
{\bm e}\cdot\bm{b} }{{\mathcal D}_2}+\frac{ {\bm e}\cdot\bm{G} }{\omega
_3{\mathcal D}_2}\Bigl[-\omega
_2\kappa _3B  \nonumber\\
&&  + 2\kappa _3\left( \kappa _2^2+\varepsilon ^2\right) \left(
{\bm{e}}^*\cdot{\bm{\theta}}_2\right) \left( {\bm e}\cdot\bm{b}\right)
+m^2\left( \omega _2\omega _3-2\varepsilon \kappa _3\right) \Bigr]\Biggr]
\Biggr\} \,,
\end{eqnarray}
where the following notation is used

\begin{eqnarray}
&&\kappa_2=\omega_2-\varepsilon\,,\quad \kappa_3=\omega_3+\varepsilon\,, \quad
\bm{\theta}_{23}=\bm{\theta}_{2}-\bm{\theta}_{3}\,,\quad A=m^2+{a}^2\,,\quad
B=m^2+{b}^2\nonumber\,,
\nonumber\\
&& \bm{a}=\bm{x}_-+\kappa_2\bm{\theta}_2\,, \quad
\bm{b}=\bm{x}_+-\kappa_3\bm{\theta}_3\,,\quad
\bm{c}=\bm{x}_+-\varepsilon\bm{\theta}_{23}\,,\quad
\bm{x}_\pm=\bm{x}\pm\bm{\Delta}_\perp/2\,,
\nonumber\\
&&
 N={8e^3Z\alpha}[\pi ^2\Delta^2\omega _1\omega _2\omega_3]^{-1}\,,\quad
\omega_1^2{\mathcal D}_1=\left(\kappa_2\bm{x}_++\kappa_3\bm{x}_-\right)^2
-\omega_2\omega_3\kappa_2\kappa_3{\theta}_{23}^2-i0\,, \nonumber\\
&&
 \omega_3^2{\mathcal D}_2=\left(\varepsilon\bm{x}_+-\kappa_3\bm{x}_-\right)^2
 -\omega_1\omega_2\kappa_3\varepsilon{\theta}_{2}^2\,,\quad
 \omega_2^2{\mathcal D}_3=\left(\kappa_2\bm{x}_++\varepsilon\bm{x}_-\right)^2
 +\omega_1\omega_3\kappa_2\varepsilon{\theta}_{3}^2\, .\nonumber
\end{eqnarray}

The functions $\bm G$ and $G_1$ are
\begin{eqnarray}
\label{G}
 {\bm G} &=&\frac{\pi(1+\xi) }{\sinh (\pi Z\alpha)} \mathrm{Im}
 \left\{\left[
 \frac{\bm{x}_+}{c_{+}}\left({c_+\over c_-}\right)^{iZ\alpha}
 \!\!\!-\frac{\bm{x}_-}{c_{-}}\left({c_-\over c_+}\right)^{iZ\alpha}
 \right] P_{iZ\alpha}^\prime(\xi)\right\}\,,\nonumber\\
 G_1 &=&\frac{\pi(1+\xi) }{\sinh (\pi Z\alpha)} \mathrm{Im}
 \left\{\left[
 \frac{c_+-2m^2}{2c_{+}}\left({c_+\over c_-}\right)^{iZ\alpha}
 \!\!\!-\frac{c_--2m^2}{2c_{-}}\left({c_-\over c_+}\right)^{iZ\alpha}
 \right] P_{iZ\alpha}^\prime(\xi)\right\}\,,
\end{eqnarray}
where $c_{\pm }=m^2+{x}_\pm^2$, $\,\xi=2m^2{\Delta}_\perp^2/
(c_{+}c_{-})-1$, and
$P_{iZ\alpha}^\prime(\xi)\,=\,dP_{iZ\alpha}(\xi)/d\xi$ is the
derivative of the Legendre function. The amplitudes with other
helicities can be obtained from (\ref{HighEnergy:Eq:final}) using
the following relations
\begin{eqnarray}
M_{+-+}(\bm{k}_1,\bm{k}_2,\bm{k}_3)&=&
M_{++-}(\bm{k}_1,\bm{k}_3,\bm{k}_2),\nonumber\\
M_{-\lambda_2\lambda_3}(\bm{k}_1,\bm{k}_2,\bm{k}_3)&=&
M_{+\Lambda_2\Lambda_3}(\bm{k}_1,\bm{k}_2,\bm{k}_3)\,
(\bm{e}\leftrightarrow \bm{e}^*) ,
\end{eqnarray}
where $\Lambda$ denotes the helicity opposite to $\lambda$.

Since the functions $\bm G$ and $G_1$ are independent of the
energy $\varepsilon$, the integrands in
(\ref{HighEnergy:Eq:final}) are rational functions of
$\varepsilon$, in which all the denominators are quadratic forms
of this variable. Therefore, the integrals over $\varepsilon$ can
be expressed via elementary functions. Resulting formulae being
rather cumbersome are not presented here explicitly. Performing
the integration over $\varepsilon$ in (\ref{HighEnergy:Eq:final}),
one obtains a twofold integral over $\bm{x}$ for the amplitudes of
photon splitting. It has been checked numerically that in the
limit $Z\alpha\rightarrow 0$ our results
(\ref{HighEnergy:Eq:final}) agree with those, obtained in
\cite{JMO80} in the Born approximation.

Let us comment on the appearance of the total momentum transfer
squared $\Delta^2$ (see (\ref{HighEnergy:Eq:Delta})) in the
coefficient $N$ in (\ref{HighEnergy:Eq:final}). The amplitudes
(\ref{HighEnergy:Eq:final}) were derived at the assumption
$\Delta_\perp\gg \Delta_z$ when there is no difference between
$\Delta^2$ and $\Delta_\perp^2$. Nevertheless, if we retain the
term $\Delta_z^2$ in $\Delta^2$ in the coefficient $N$, the
formulas (\ref{HighEnergy:Eq:final}) become valid also for
$\Delta_\perp\sim \Delta_z$. This can be explained as follows. In
the region $\Delta_\perp\sim \Delta_z$ the Born amplitude prevails
over the Coulomb corrections (cf. (\ref{AppendixC:Eq:zero}) and
(\ref{AppendixC:Eq:CCzero})). On the other hand, in the
small-angle approximation it contains the momentum transfer in the
form $\bm{\Delta}_\perp/\Delta^2$, where $\Delta$ is just the
total momentum transfer.

For $\omega_2\theta_2,\,\omega_3\theta_3\gg m$ we can neglect the
electron mass in the expressions (\ref{HighEnergy:Eq:final}). In
this case (zero mass limit) the amplitudes are simplified, which
allows us to perform the further analytical investigation  of the
process (see \ref{AppendixC}).

\subsection{Cross section}

\begin{figure}[h]
  \centering
  \includegraphics[height=220pt,keepaspectratio=true]{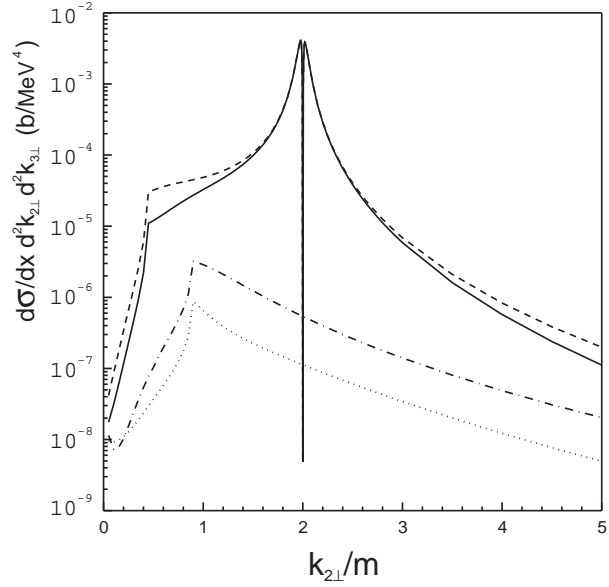}
  \caption{Differential cross section
  $d\sigma/dxd\bm{k}_{2\perp}d\bm{k}_{3\perp}$ vs ${k}_{2\perp}/m$
  in a screened Coulomb potential for different azimuth angle $\phi$ between
  vectors $\bm{k}_{2\perp}$ and $\bm{k}_{3\perp}$; $Z=83$, $x=0.1$,
$\omega_1=1$GeV, $k_{3\perp}=2m$.  The dashed curve (Born approximation) and
the solid curve (exact cross section) correspond to $\phi=\pi$. The dash-dotted
curve (Born approximation) and the dotted curve (exact cross section)
correspond to $\phi=0$.}
  \label{HighEnergy:Fig:DifDif}
\end{figure}

In the small-angle approximation ($\theta_{2,3}\ll 1$) the cross
section (\ref{General:CrossSection}) can be represented in the
form
\begin{equation} \label{cross1}
d\sigma=|M|^2\,\frac{dx \, d\bm{k}_{2\perp}\, d\bm{k}_{3\perp}\, }{2^8
\pi^5\omega_1^2x(1-x)} \,.
\end{equation}
In this subsection a screened Coulomb potential is used in the
numerical calculations. As was explained above, to take screening
into account, the Born part of the amplitude should be multiplied
by an atomic form factor, e.g. (\ref{LowEnergy:FF}). To illustrate
a magnitude of the Coulomb corrections, the exact and Born
differential cross sections for unpolarized photons
$d\sigma/dxd\bm{k}_{2\perp}d\bm{k}_{3\perp}$ are plotted in Fig.
\ref{HighEnergy:Fig:DifDif} as a function of $k_{2\perp}/m$ for
${k}_{3\perp}=2m$, $x=0.1$, $\omega_1=1$GeV and the azimuth angle
(the angle between the vectors $\bm{k}_{2\perp}$ and
$\bm{k}_{3\perp}$) $\phi=0$ and $\pi$. The value  $Z=83$ (bismuth)
was chosen since bismuth atoms determine the cross section of
photon splitting in the experiment \cite{Akh97}.
A wide peak for azimuth angle $\phi=\pi$ is due to small momentum
transfer $\Delta$. There is a narrow notch (c.f. Fig.
\ref{LowEnergy:SigDif(theta)})  in the middle of this peak ( at
$k_{2\perp}/m=2$ ) where the condition
$\bm{\Delta}_\perp=\bm{k}_{2\perp}+\bm{k}_{3\perp}=0$ is
fulfilled. Such a structure was discussed first in \cite{St83}.
The width of the notch is about
$\mathrm{max}(\Delta_z/m,\beta_0/m)$. Recall that $\Delta_z$ is
the longitudinal component of the momentum transfer defined by
(\ref{HighEnergy:Eq:Delta}), and $\beta_0=m Z^{1/3}/121$
(\ref{LowEnergy:FF}) characterizes the effect of screening. In our
example $\beta_0$ is larger than $\Delta_z$ , so the width of the
notch is roughly $\beta_0/m=3.6\times 10^{-2}$. Let us note that
for $\omega_1\gg m$ the differential cross section, expressed in
terms of $\bm{k}_{2\perp}$, $\bm{k}_{3\perp}$, $x$, and
$\omega_1$, depends on $\omega_1$ only via $\Delta_z$. Due to this
fact, the differential cross section is independent of $\omega_1$
in the region where we can neglect $\Delta_z$, i.e. outside the
notch vicinity. For small $\Delta$ as compared to $k_\perp$ the
Born amplitude, which is proportional to $1/\Delta$, is much
larger than the Coulomb corrections. That is why the exact and
Born cross sections coincide within the peak region. Outside this
region the Coulomb corrections essentially modify the cross
section. The points of discontinuous slope of the curves in Fig.
\ref{HighEnergy:Fig:DifDif} are related to the threshold
conditions for the production of real electron-positron pair by
two photons with the momenta $\bm{k}_2$ and $\bm{k}_3$:

\begin{equation}\label{HighEnergy:Eq:threshold}
(k_2+k_3)^2 =\omega_2\omega_3\theta_{23}^2=4m^2\ .
\end{equation}

It is worth noting that the cross section
$d\sigma_{\lambda_1\lambda_2\lambda_3}/dx\,d\Omega_2d\Omega_3$ for
definite helicities of photons is not an even function of the
azimuth angle $\phi$. Instead, due to the parity conservation the
following relation holds

\begin{equation}
\frac{d\sigma_{\lambda_1\lambda_2\lambda_3}}{dx\,d\Omega_2d\Omega_3}
(\phi)=\frac{d\sigma_{\Lambda_1\Lambda_2\Lambda_3}}{dx\,d\Omega_2d\Omega_3}
(-\phi)
\end{equation}

\begin{figure}[h]
  \centering
  \includegraphics[height=130pt,keepaspectratio=true]{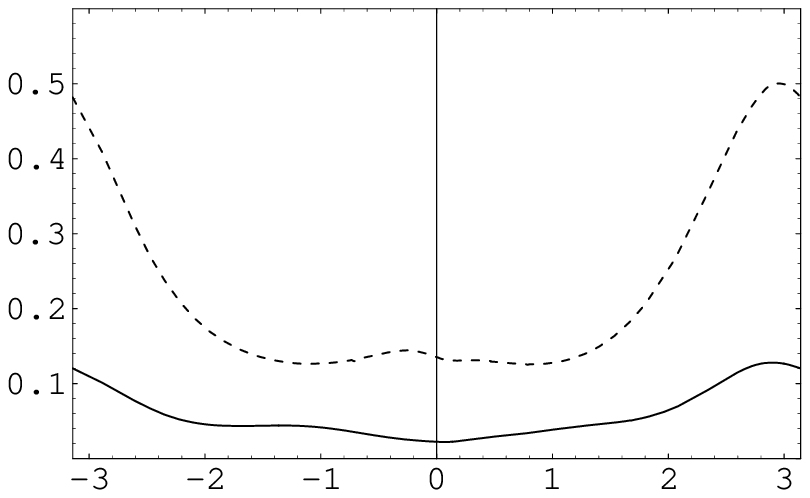}
  \includegraphics[height=133pt,keepaspectratio=true]{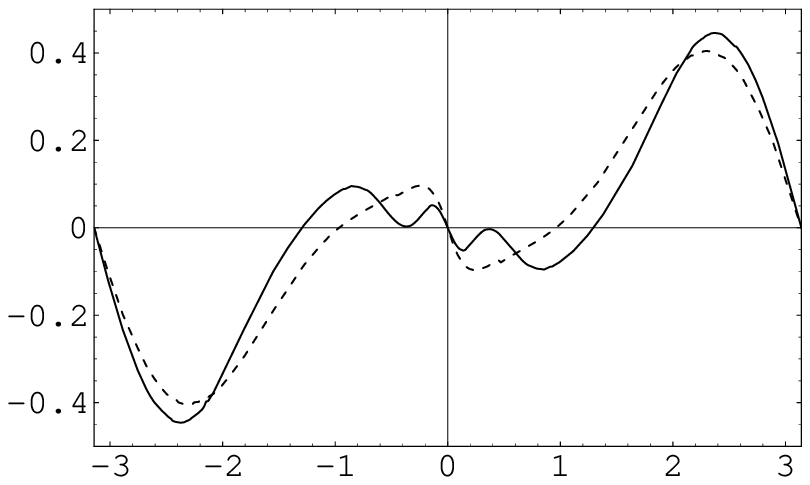}
 \begin{picture}(0,0)(0,0)
 \put(-405,115){ $S_+$}
 \put(-310,115){$a$}
 \put(-325,-7){ $\phi$}
 \put(-180,115){ $\xi$}
 \put(-95,115){$b$}
 \put(-105,-7){ $\phi$}
 \end{picture}
  \caption{The dependence of $S_+$ ($a$) and $\xi=2(S_+-S_-)/(S_++S_-)$
 ($b$) on the azimuth angle $\phi$ at $Z=83$, $\omega_1=1$GeV, $x=0.3$,
 $\theta_2=0.014$, and $\theta_3=0.02$. Here $S_\pm=\sum_{\lambda_2\lambda_3}
d\sigma_{\pm\lambda_2\lambda_3}/dx\,d\Omega_2d\Omega_3$ are the
differential cross sections for positive ($S_+$) and negative
($S_-$) helicities of the initial photon. Solid curves represent
the exact in $Z\alpha$ results, dashed curves are obtained in the
Born approximation.} \label{HighEnergy:Fig:Asym}
\end{figure}

As an illustration of the azimuth asymmetry of the cross section,
the dependence of $S_+$ and $\xi=2(S_+-S_-)/(S_++S_-)$ on the
azimuth angle $\phi$ is shown in Fig. \ref{HighEnergy:Fig:Asym},
where $S_\pm=\sum_{\lambda_2\lambda_3}
d\sigma_{\pm\lambda_2\lambda_3}/dx\,d\Omega_2d\Omega_3$ are the
differential cross sections for positive ($S_+$) and negative
($S_-$) helicities of the initial photon. Again, the exact in
$Z\alpha$ cross section (solid curve in Fig.
\ref{HighEnergy:Fig:Asym}$a$) is much smaller than the cross
section obtained in the Born approximation (dashed curve
\textit{ibid}). The magnitude of azimuth asymmetry (Fig.
\ref{HighEnergy:Fig:Asym}$b$) reaches tens of percent. For the
parameters used in our example the effect of screening is not
essential.

\begin{figure}[h]
  \centering
  \includegraphics[height=220pt,keepaspectratio=true]{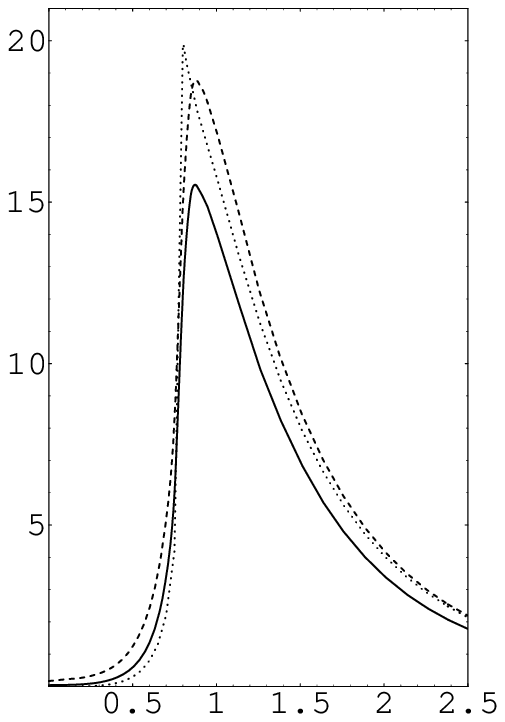}
  \includegraphics[height=220pt,keepaspectratio=true]{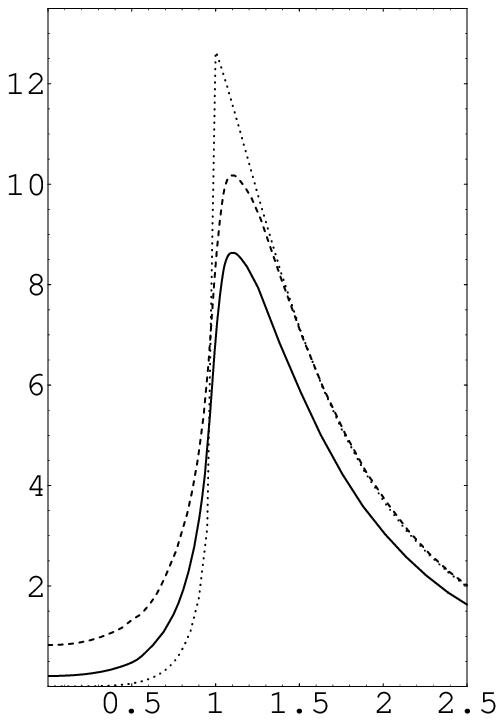}
  \includegraphics[height=220pt,keepaspectratio=true]{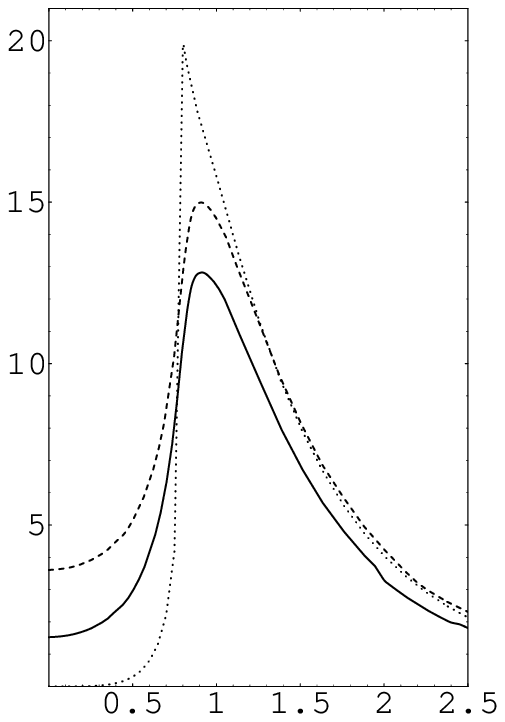}
 \begin{picture}(0,0)(0,0)
 \put(-225,2){ $k_{2\perp}/m$}
 \put(-315,195){ $a$}
 \put(-175,195){ $b$}
 \put(-35,195){ $c$}
 \put(-435,85){\rotatebox{90}{$m^2\sigma_0^{-1} d\sigma/dxd\bm{k}_{2\perp}$}}
 \end{picture}
  \caption{$m^2\sigma_0^{-1} d\sigma/dxd\bm{k}_{2\perp}$ vs $k_{2\perp}/m$
for a screened Coulomb potential, $\omega_1/m=1000$, $Z=83$,
$\sigma_0=\alpha^3(Z\alpha)^2/(4\pi^2m^2)=0.782\cdot 10^{-9} \,
Z^2 \mathrm{ b}$, $x=0.2$ ($a$), $0.5$ ($b$), and $0.8$ ($c$). The
dashed curves correspond to the Born approximation, the solid
curves give the exact result, and the dotted curves show the
results of the WW approximation.}
  \label{HighEnergy:Fig:Dif}
\end{figure}

In Fig. \ref{HighEnergy:Fig:Dif} the differential cross section $m^2
\sigma_0^{-1} d\sigma/dx\, d\bm{k}_{2\perp}$ is shown depending on
$k_{2\perp}/m$ for a screened Coulomb potential at $\omega_1/m=1000$, $Z=83$,
and different $x$,
$$
\sigma_0=\frac{\alpha^3(Z\alpha)^2}{4\pi^2m^2}=0.782\cdot 10^{-9} \, Z^2 \mbox{
b}\ .
$$
Solid curves represent the exact cross sections, the dashed curves
give the Born results, and the dotted curves correspond to the WW
approximation (\ref{WW:CS(theta)}). The cross section exhibits a
thresholdlike behavior in the vicinity of the point
$k_{2\perp}=k_{th}=2\sqrt{x(1-x)}m$, where both conditions
$\bm{\Delta}_\perp=\bm{k}_{2\perp}+\bm{k}_{3\perp}=0$ and
(\ref{HighEnergy:Eq:threshold}) hold. Under these conditions the
peak in the cross section
$d\sigma/dxd\bm{k}_{2\perp}d\bm{k}_{3\perp}$ seats on the boundary
of the kinematic region in which the production of a real
electron-positron pair by two photons with the momenta $\bm{k}_2$
and $\bm{k}_3$ is possible. The cross section $d\sigma/dx\,
d\bm{k}_{2\perp}$ drops rapidly for $k_{2\perp}\gg m$ ($\propto
1/k_{2\perp}^4$). As seen in Fig. \ref{HighEnergy:Fig:Dif}, the
Coulomb corrections to the cross section integrated over
$\bm{k}_{3\perp}$ noticeably diminish the magnitude of the cross
section. Above the threshold ($k_{2\perp}>k_{th}$) the Coulomb
corrections reach tens of percent while below the threshold the
exact cross section is several times smaller than the Born one. It
can be explained as follows.  Above the threshold the main
contribution to the cross section $d\sigma/dx\, d\bm{k}_{2\perp}$
is given by the integration region with respect to
$\bm{k}_{3\perp}$ where
$\mbox{max}(\bm{k}_{2\perp}^2/\omega_1,\beta_0)\ll\Delta\ll
k_{2\perp}$. As a result, the Born cross section is
logarithmically amplified as compared to the Coulomb corrections.
Far below the threshold where $k_{2\perp}\ll k_{th}$, the region
$\Delta\ll k_{2\perp}$ does not contribute to the cross section
$d\sigma/dx\, d\bm{k}_{2\perp}$ since in this region
$k_{3\perp}\approx k_{2\perp} \ll m$, and the amplitude is
suppressed as a power of $k_{2\perp}^2/m^2$. Therefore, below the
threshold the main contribution to the cross section is given by
the region $k_{3\perp}\sim m$, where the amplitude, exact in
$Z\alpha$, drastically differs from the Born one. Due to the same
reason below the threshold the cross section in the WW
approximation (dotted curves) differs essentially from the Born
result because the former comes from the region $\Delta\ll
k_{2\perp}$. As follows from (\ref{AppendixWW:Eq:MA}), for
$k_\perp\ll m$ the amplitudes entering the cross section
(\ref{WW:CS(theta)}) are suppressed as $k_\perp^4/m^4$. One can
see from Fig. \ref{HighEnergy:Fig:Dif} that in accordance with the
expression for $k_{th}$ the position of the peak is the same for
$x=0.2$ and $x=0.8$. Nevertheless, the magnitude  of these two
cross sections are significantly different, especially below the
threshold. However, as expected and verified numerically, the
cross section $d\sigma/dx$ at $x=0.2$ coincide with that at
$x=0.8$.

\begin{figure}[h]
  \centering
  \includegraphics[height=180pt,keepaspectratio=true]{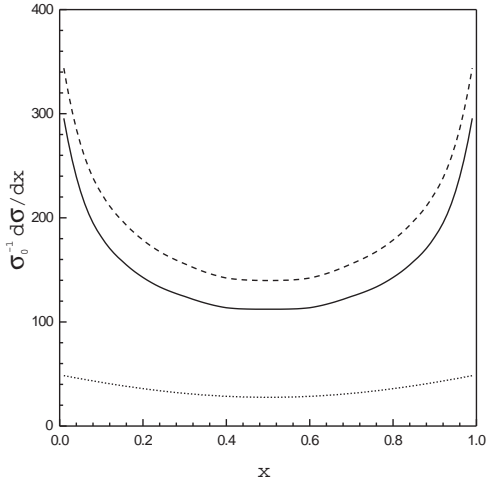}
  \caption{The dependence of $\sigma_0^{-1}d\sigma/dx$ on $x$ for a
screened Coulomb potential, $\omega_1/m=1000$, $Z=83$.  The dashed curve
corresponds to the Born approximation, the solid one gives the exact result,
and the dotted curve shows the difference between the Born cross section and
the exact one.}
  \label{HighEnergy:Fig:Spectr}
\end{figure}

 Let us consider now the magnitude of the Coulomb corrections to the
cross section  $d\sigma/dx$ integrated over the transverse momenta of both
final photons. The main contribution to this cross section is given by the
region where ${k}_{2\perp},\,{k}_{3\perp} \sim m$.
 The Born contribution to $d\sigma/dx$ contains
large logarithm resulting from the integration over small momentum
transfer region $\mbox{max}(\beta_0,m^2/\omega_1)\ll\Delta\ll m$.
For $\beta_0\gg m^2/\omega_1$, the cross section $d\sigma/dx$ is
independent of $\omega_1$, while for $\beta_0\ll m^2/\omega_1$ it
grows slowly (as $\ln (\omega_1/m)$) when $\omega_1$ increases.
Since the Coulomb corrections to $d\sigma/dx$ are determined by
the region of momentum transfer $\Delta\sim m$, they do not depend
on $\omega_1$ for $\omega_1\gg m$ . They also are insensitive to
the effect of screening. In Fig. \ref{HighEnergy:Fig:Spectr} the
exact (solid curve) and the Born (dashed curve) cross sections
$\sigma_0^{-1} d\sigma/dx$ are plotted as functions of $x$ at
$\omega_1/m=1000$, and $Z=83$. As should be, the curves are
symmetric with respect to the replacement $x\rightarrow 1-x$. The
dotted curve shows the Coulomb corrections (the difference between
the exact in $Z\alpha$ cross section and the Born one) taken with
the opposite sign. Note that their dependence on $x$ is rather
weak. The cross section $d\sigma/dx$ in Fig.
\ref{HighEnergy:Fig:Spectr} increases rapidly when $x$ approaches
the points $x=0$ and $x=1$. However, this cross section should
vanish at $x=0$ and $x=1$ due to the gauge invariance of QED. It
turns out that the decrease of the cross section $d\sigma/dx$
occurs in the very narrow interval $\delta x\sim m^2/\omega_1^2$
close to the points $x=0,1$. Therefore, the contribution from
these domains to the total cross section $\sigma$ is negligible.
In our example ( $Z=83$, $\omega_1/m=1000$ ), the exact result for
$\sigma$ is $3.9\times 10^{-4}$b while the Born approximation
gives $4.8\times 10^{-4}$b, the difference being 23\%.

\begin{figure}[h]
\centering
\includegraphics[height=180pt,keepaspectratio=true]{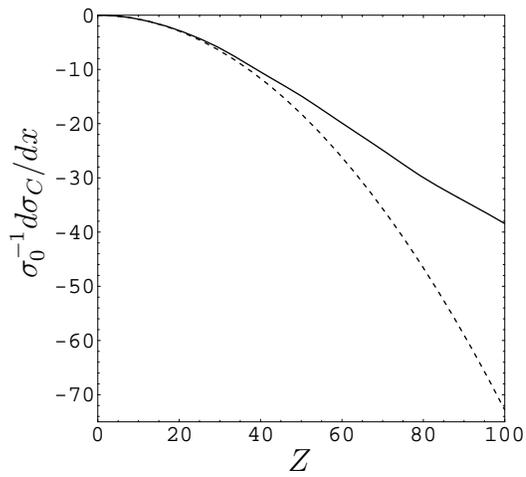}
\put(-90,-3){$Z$} \put(-195,70){\rotatebox{90}{$\sigma_0^{-1}
d\sigma_C/dx$}} \caption{The dependence of the Coulomb corrections
$\sigma_0^{-1} d\sigma_C/dx$ on $Z$ for $x=0.7$; solid curve:
complete result, dashed curve: first correction.}
  \label{HighEnergy:Fig:CC}
\end{figure}

In Fig. \ref{HighEnergy:Fig:CC} the Coulomb corrections to the
cross section $d\sigma_c/dx$ divided by $\sigma_0=0.782\cdot
10^{-9} Z^2$b are shown as a function of $Z$ for $x=0.7$ (solid
line). Since their dependence on $x$ is rather weak ( see Fig.
\ref{HighEnergy:Fig:Spectr}), this curve allows one to estimate
the magnitude of the Coulomb corrections for any $x$. To elucidate
the role of higher-order Coulomb corrections we plot also the
first Coulomb correction to the cross section divided by
$\sigma_0$ (dashed curve). This ratio is proportional to $Z^2$. As
seen in Fig. \ref{HighEnergy:Fig:Spectr}, for sufficiently large
$Z$ the first Coulomb correction is irrelevant to the complete
result for these corrections.

To conclude this Section, we emphasize that the proper description
of photon splitting in the electric field of a heavy atom requires
the exact account of this field. For large $Z$, the contribution
of the Coulomb corrections is always noticeable, though the
magnitude of the effect depends on the type of a cross section and
kinematic conditions.

\section{Experimental investigations of photon splitting}

The observation of photon splitting is a difficult problem due to
a smallness of its cross section   as compared to those of other
processes initiated by the initial  photons  in a target. The
following background processes are significant:  double Compton
effect off the atomic electrons ($\gamma e\to \gamma\gamma e$),
and the emission  of two hard photons from  $e^+e^-$ pair produced
by the initial photon. The relative importance of these processes
depends on the photon energy. For the energy  $\omega_1 \sim m$,
where the search of photon splitting was undertaken in two
experiments \cite{Adler66,Rob66}, only double Compton scattering
determines the background conditions. In these experiments,  the
photons from an intense radioactive source ($Zn^{65}$ with
$\omega_1~\simeq~1.1$~MeV in \cite{Adler66}, and $Co^{60}$ with
$\omega_1~\simeq~1.17,\, 1.33$~MeV in \cite{Rob66}) were used. The
combination of the coincidence and energy-summing detection
technique was applied. The number of events considered as
candidates for photon splitting exceeded the theoretical
expectations by the  factor of~300 in~\cite{Rob66}, and by the
factor of~6 in~\cite{Adler66}. The role of the background process
in \cite{Adler66} was unclear. Nevertheless, the authors
of~\cite{Adler66} concluded that their "results for $Z\sim28$
strongly indicate the existence of photon splitting".

At high photon energy $\omega\gg m$ the emission  of hard photons
from  $e^+e^-$ pair becomes  most important as a background
process. In 1973 the experiment devoted to the study of Delbr\"uck
scattering of photons in energy region $1\div 7$~GeV was performed
\cite{DESY}. The bremsstrahlung non-tagged photon beam was used.
Some events were assigned by authors of~\cite{DESY} to the photon
splitting process. As shown in \cite{BKKF,dzhilk},  the
theoretical value for the number of photon splitting events under
the conditions of the experiment was two orders of magnitude
smaller than the experimental result. It was also argued that the
events observed could be explained by the production of $e^+e^-$
pair and one hard photon.

The first successful observation of photon splitting  was
performed in 1995-96 using the tagged photon beam of the energy
$120\div450$~MeV at the VEPP-4M $e^+e^-$ collider~\cite{vepp} in
the Budker Institute of Nuclear Physics (Novosibirsk). Another
goal of this experiment was a study of Delbr\"uck
scattering~\cite{Akh98}. The total statistics collected
was~$1.6\cdot10^9$ incoming photons with~BGO (bismuth germanate)
target and $4\cdot10^8$ photons without target for background
measurements. The preliminary results were published in
\cite{Akh97}, \cite{ph98}. Here we present the last  data analysis
for this experiment \cite{ph2001}.

\begin{figure}[htb]
\centering
\includegraphics*[width=0.4\textwidth]{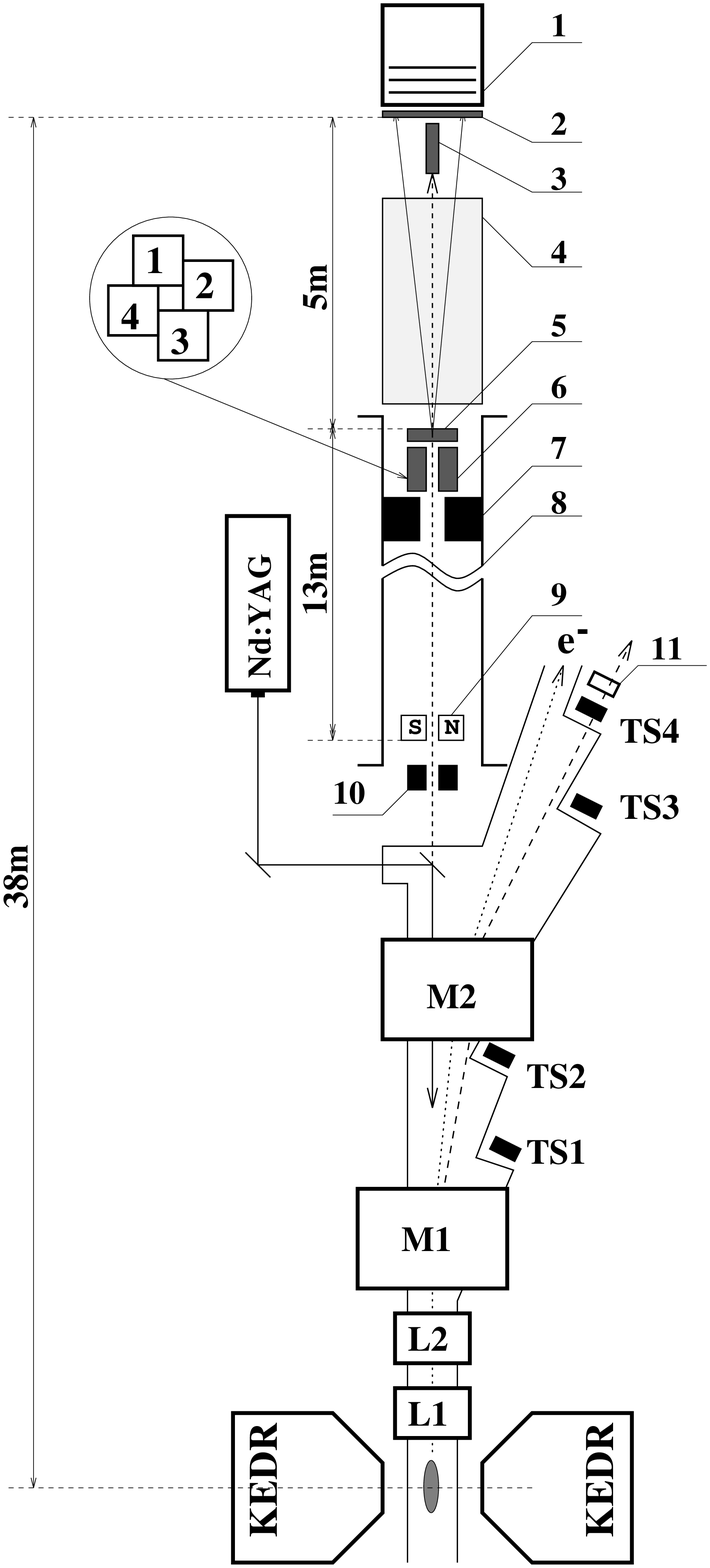}
\caption{Principal scheme of setup: LKr calorimeter (1);
scintillation veto-counter (2); beam  dump (BGO) (3); He-filled
tube (4m length) (4); target (BGO) (5); active collimator (BGO)
(6); lead absorber (7); guiding tube for the gamma-quanta beam
(8); cleaning magnet (9); passive lead collimator (10); TS
scintillattion counter (11); Nd:YAG is the laser; TS1-TS4 are
tagging system hodoscopes; M1 and M2 are bending magnets; L1 and
L2 are quadrupole lenses.} \label{Experiment:Fig:ps-setup}
\end{figure}

 The experimental setup is
shown in Fig.~\ref{Experiment:Fig:ps-setup}. Some ideas of this
setup were suggested in~\cite{MW91}. The main features of the
experimental approach are:
\begin{itemize}
\item The use of high-quality tagged photon beam produced by
backward Compton scattering of laser light off high-energy
electron beam. Thereby, the energy of the initial photon is
accurately determined.
\item Strong suppression of the background processes by means of the detection
 of charged particles produced in the target and in other elements
 of the photon-beam line.
\item The detection of both final photons to discriminate the
photon splitting events from those with one final photon produced
in Compton or Delbr\"uck scattering.
\item The requirement  of the balance  between the sum of the  energies of the
final photons and the energy of the tagged initial photon. This
provides the  additional suppression of the events with charged
particles missed by the detection system.

\end{itemize}

At high energy of the initial photon $\omega_1 \gg m$, the photon
splitting cross section is peaked at small angles between momenta
of all photons ($\sim m / \omega_1$). Therefore, a good
collimation of the primary photon beam is required. The ROKK-1M
facility~\cite{ROKK} is used as the intense source of the tagged
$\gamma$-quanta. The electron energy loss in the process of
Compton scattering of laser light is measured by the tagging
system (TS)~\cite{TS} of the KEDR detector~\cite{kedr}. The TS
consists of the focussing spectrometer formed by accelerator
quadrupole lenses and bending magnets, and 4 hodoscopes  of the
drift tubes. High-energy photons move in a narrow cone around the
electron beam direction. The angular spread is of the order
$1/\gamma$, where $\gamma=E_{beam}/m$ is the relativistic factor
of the electron beam. The photon energy spectrum has a sharp edge
at
\begin{equation}
   \omega_{th}=\frac{4\gamma^2\omega_{laser}}{1+4\gamma\omega_{laser}/m}\,,
\end{equation}
that allows one to perform the absolute calibration of the tagging
system in a wide energy range. In experiment the laser photon
energy was $\omega_{laser}=1.165$~eV, the electron beam energy
$E_{beam}=5.25$~GeV, and $\omega_{th}=450$~MeV. The photon energy
resolution provided by the tagging system depends on the photon
energy and on the position of scattered electron in TS hodoscope:
it was 0.8~\% at $\omega_1=450$~MeV (at the center of the
hodoscope) and $\sim5$~\% at $\omega_1=120$~MeV (at the edge of
the hodoscope).
     The collimation of the photon beam is provided by two collimators
spaced at 13.5~m. The last collimator, intended to strip off the
beam halo produced on the first one, is made of four BGO (bismuth
germanate) crystals as shown in the separate view in
Fig.~\ref{Experiment:Fig:ps-setup}. After passing through the
collimation system, the photon beam  hits $1X_0$ thick BGO crystal
target.

Since one has to separate incident photons from those scattered in
the target, certain angular region around the photon beam
direction $\theta \le \theta_0$ is enclosed by the dump made of
13~$X_0$ thick BGO crystal  installed in front of the photon
detector. The incoming photons which passed the target without
interaction are absorbed by the dump, and the only photons to be
detected are those scattered outside the dump shadow. For too
large value of $\theta_0$,  many photon splitting events  would be
lost while for too small value of $\theta_0$, the non-interacted
photons would overwhelm  the data acquisition system. As an
optimum, the value $\theta_0=2.4$~mrad was chosen.

 All active elements used in
beam line (collimators, target, dump, scintillating veto counter)
set a veto signal in the trigger and their signals are  used in
the analysis for background suppression. The information from the
target and beam dump is also used for measurement of the incoming
photon flux. The liquid-krypton ionization calorimeter is used for
the detection of the final photons.  Its three-layer double-sided
electrode structure enables one to get both (X and Y) coordinates
for detected photons. The energy resolution of the calorimeter
is~$2\,\% / \sqrt{\omega(\mbox{GeV})}$. The liquid-krypton
calorimeter is described in details in~\cite{prot,lkrpos}.

\begin{figure}[htb]
\centering
\includegraphics*[width=0.9\textwidth]{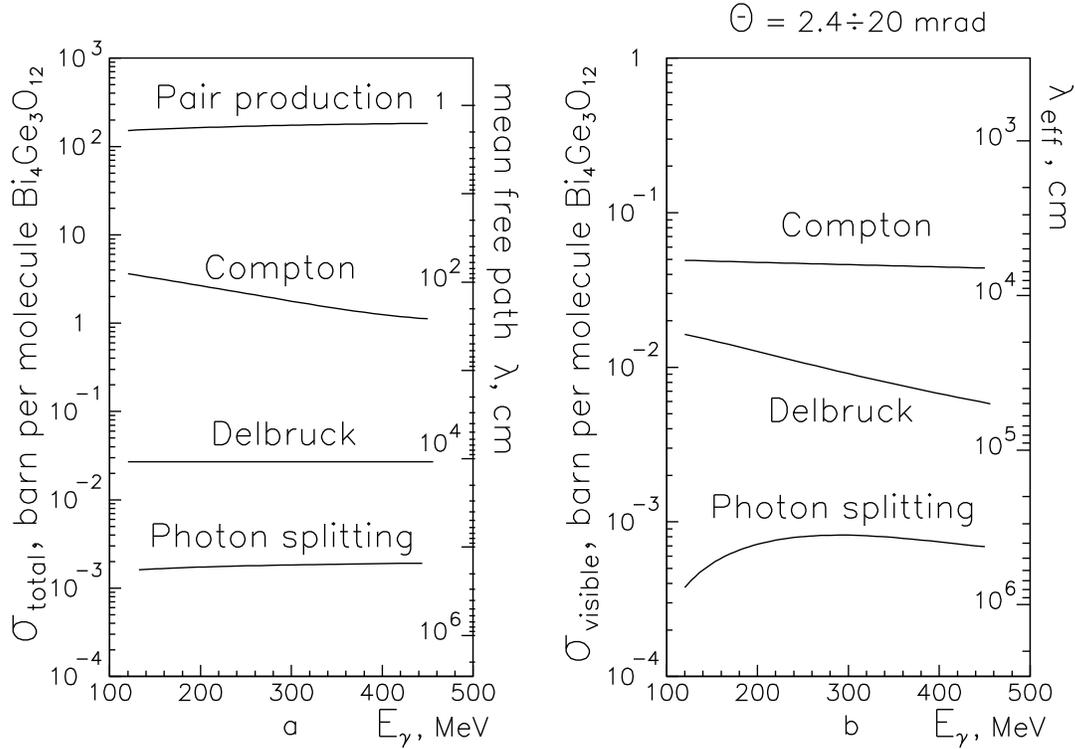}
\caption{The calculated energy dependence of the total (a) and the
visible (b) cross sections of various processes initiated by
photons in BGO~target (in units of barn per one molecule
of~$Bi_{4}Ge_{3}O_{12}$).} \label{Experiment:Fig:crsec}
\end{figure}

In the experiment, the detected final photons had the polar angles
in the region $2.4~\mbox{mrad}\le \theta \le 20~\mbox{mrad}$. The
corresponding cross section is called  "visible".
Fig.~\ref{Experiment:Fig:crsec} shows the calculated energy
dependence of the total ($a$) and visible ($b$) cross sections for
various processes initiated by photons in BGO~target: $e^+e^-$
pair production, Compton scattering on atomic electrons,
Delbr\"uck scattering, and photon splitting. The calculation of
the photon splitting cross section was performed using the results
obtained in \cite{LMS1,LMS2,LMS3}.

In the event selection procedure the following constraints   were
applied:
\begin{itemize}
\item The absence of the signal caused by charge particles in all active
elements of the photon-beam line.
\item The  balance of  the tagged initial
photon energy and the energy measured in the calorimeter within
$3\sigma$ of its energy resolution.
\item The existence of two separate tracks at least for one (X or Y) coordinate
 in the calorimeter strip structure. The track is found if there
 are close clusters in different layers.
\end{itemize}

The fulfillment of the latter  requirement strongly suppresses the
contribution of the processes with one photon in the final state
which could imitate two photon events in the calorimeter.

\begin{figure}[htb]
\centering
\includegraphics*[width=0.8\textwidth]{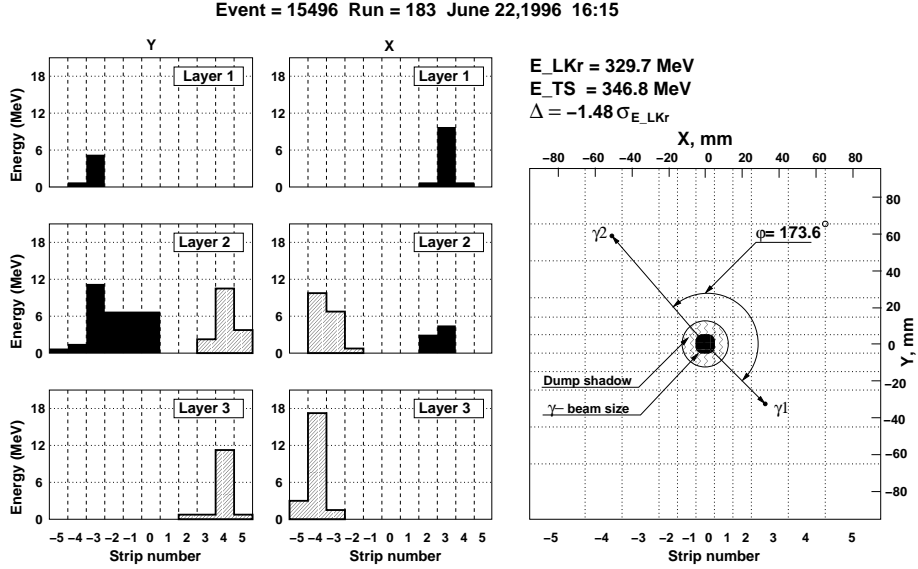}
\caption{Energy profile in the calorimeter strip structure (left)
and reconstructed kinematics (right) for a typical candidate to
the photon splitting event.} \label{Experiment:Fig:event}
\end{figure}

The typical event which meets selection criteria is shown in
Fig.~\ref{Experiment:Fig:event}. In this example two tracks are
seen in both X and Y directions. The conversion of the first
photon  occurs in the Layer 1 while the second photon converts in
the Layer 2.

\begin{figure}[h]
\centering
\includegraphics*[width=0.5\textwidth]{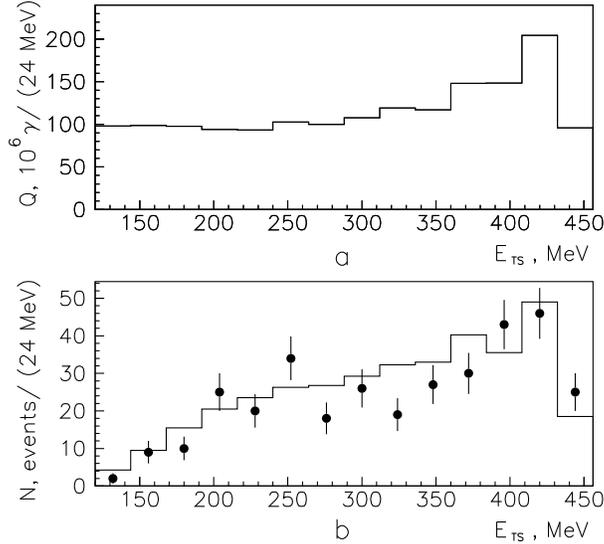}
\caption{(a) The photon energy spectrum  measured in the tagging
system (TS). (b) The number of reconstructed photon splitting
events as a function of the tagged photon energy $E_{TS}$. In plot
(b) black circles present the  experimental results, histogram is
the result of Monte-Carlo simulation based on the exact in
$Z\alpha $ photon splitting cross section.}
\label{Experiment:Fig:expspec}
\end{figure}

\begin{figure}[htb]
\centering
\includegraphics*[width=0.8\textwidth]{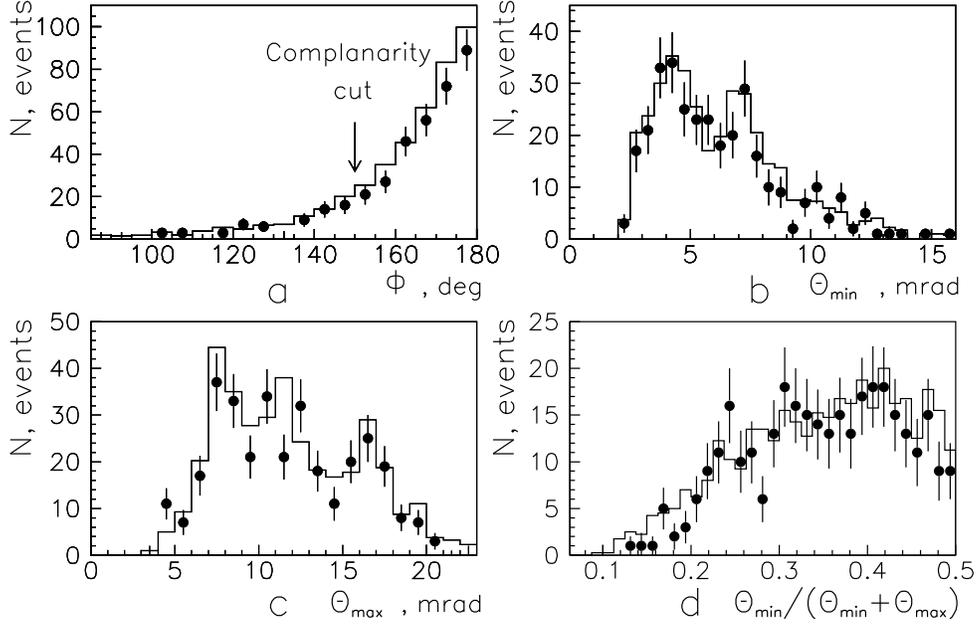}
\caption{The number  of the selected photon splitting events as a
function of
\newline
the azimuth angle between momenta of the outgoing photons (a);
\newline
 the polar angle $\theta_{min}=\mbox{min}\{\theta_2,\theta_3\}$
(b);
\newline
 the polar angle $\theta_{max}=\mbox{max}\{\theta_2
 ,\theta_3\}$
(c);
\newline
 the variable $\tilde
x={\theta_{min}}/({\theta_{min}+\theta_{max}})$ (d).
\newline
 In figures
(b), (c), and (d) only events satisfying the complanarity cut
$\phi \ge 150^\circ $ (see plot (a)) are included. Black circles
present the  experimental results, histograms are the results of
Monte-Carlo simulation based on the exact in $Z\alpha $ photon
splitting cross section.} \label{Experiment:Fig:pspic98}
\end{figure}

\newpage
The experimental results are presented in Tab.~\ref{tab:res} and
in Figs.~\ref{Experiment:Fig:expspec},
\ref{Experiment:Fig:pspic98} together with the results of
Monte-Carlo simulation based on the exact photon splitting cross
section. The energy spectrum of the initial photons measured in
the tagging system is shown in
Fig.\ref{Experiment:Fig:expspec}(a). Tab.~\ref{tab:res} and
Fig.~\ref{Experiment:Fig:pspic98} present the data summed up  over
the initial photon spectrum. The errors shown in the
Tab.~\ref{tab:res} are statistical ones. The systematic error is
determined by the accuracy  of the measurement of the number of
initial photons  and by the uncertainty in the reconstruction
efficiency of photon splitting events. The estimation of these
systematic errors gives 2~\% and 5~\%, respectively.

\begin{table}[h]
\centering  \caption{The number of reconstructed events. Here Q is
the number of incoming  photons. The quantity $N_{\phi<150^\circ
}$ is the number of events with the complanarity
angle~${\phi<150^\circ }$ (see Fig.~\ref{Experiment:Fig:pspic98}),
$N_{\phi>150^\circ }$ is the number of events with
${\phi>150^\circ }$. The quantities $N_{\phi<150^\circ}$ and
$N_{\phi>150^\circ}$ are normalized to the experimental statistics
collected with the target. MC means Monte-Carlo simulation.}
\vspace{0.5cm}
\begin{tabular}{|c|c|c|c|c|}
\hline {\bf DATA}&{\bf TARGET}&${\bf Q,\,{10^{9}}}$
&$\bf N_{\phi>150^\circ }$ & $\bf N_{\phi<150^\circ }$ \\
\hline Experiment  &$\rm Bi_{4} Ge_{3} O_{12}$ &1.63 & 336$\pm$18
& 82$\pm$9 \\
Experiment   &no target               &0.37 &10$\pm$3
&10$\pm$3 \\
MC photon splitting &$\rm Bi_{4} Ge_{3} O_{12}$ &6.52 &364$\pm$10
& 72$\pm$5  \\
MC Delbr\"uck scattering &$\rm Bi_{4} Ge_{3} O_{12}$ &1.63
&2$\pm$1
& 16$\pm$4  \\
MC other backgrounds&$\rm Bi_{4} Ge_{3} O_{12}$ &1.63 &0
& 16$\pm$4  \\
\hline
\end{tabular}
\label{tab:res}
\end{table}

As seen from the Table \ref{tab:res} and
Fig.~\ref{Experiment:Fig:pspic98}a, the main part of the photon
splitting events has, in agreement with the theory, a complanarity
angle $\phi$ (the azimuth angle between final photon momenta)
close to $180^\circ$. The choice of the interval $\phi>150^\circ$
allows us to improve the signal-to-background ratio (see, e.g.,
the last two rows of the Table \ref{tab:res}). Just this
$\phi$-interval was used to plot the distributions over polar
angles in Fig.~\ref{Experiment:Fig:pspic98} and the dependence of
the number of reconstructed photon splitting events on the initial
photon energy $E_{TS}$ in Fig.~\ref {Experiment:Fig:expspec}(b).
Note that for most of the events in this $\phi$-interval, the
variable
 $\tilde x={\theta_{min}}/({\theta_{min}+\theta_{max}})$ is
 approximately equal to the ratio
 $\mbox{min}(\omega_2,\,\omega_3)/\omega_1$ since the main
 contribution to the cross section comes from the region
 $|\bm{k}_{2\perp}+\bm{k}_{3\perp}|\ll {k}_{2\perp},\,{k}_{3\perp}$,
  i.e. $\phi\approx 180^\circ$ and
  $\omega_2\theta_2\approx\omega_3\theta_3$.

The results presented in the Table~\ref{tab:res} and in
Figs.~\ref{Experiment:Fig:expspec}(b),
\ref{Experiment:Fig:pspic98} show a good agreement between the
theory and the experiment. More precisely, the total number of
reconstructed  events in the experiment (see Table~\ref{tab:res})
differs from the result of MC simulation by 1.5 standard
deviations.

\begin{figure}[h]
\centering
\includegraphics*[width=0.5\textwidth]{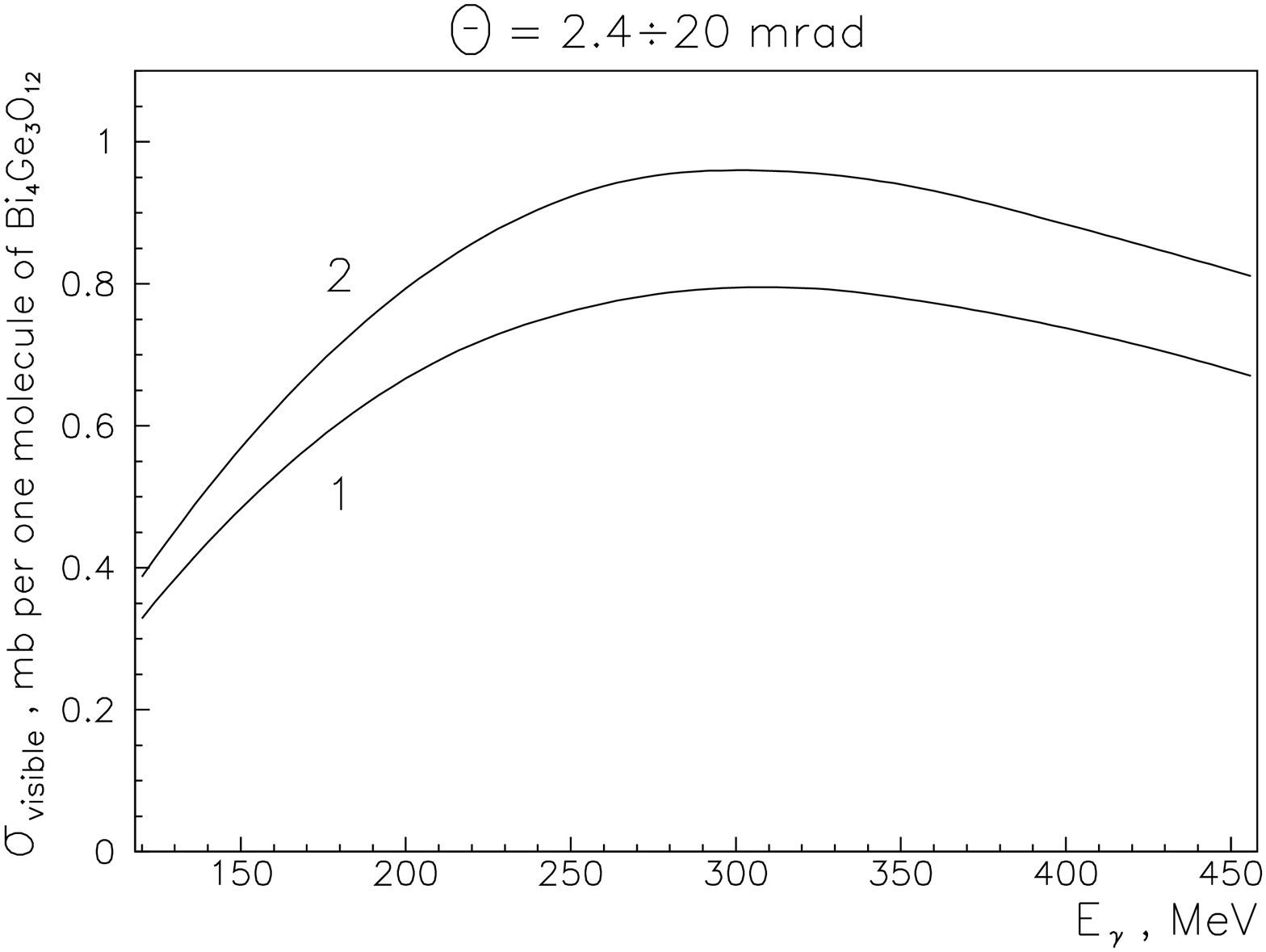}
\caption{The visible photon splitting cross section calculated
exactly in $Z\alpha$ (1) and in the Born approximation(2) as a
function of the initial photon energy.}
\label{Experiment:Fig:BornCoulomb}
\end{figure}

In order to demonstrate the role of the Coulomb corrections under
the experimental conditions, we show in
Fig.~\ref{Experiment:Fig:BornCoulomb} the visible  cross sections
calculated exactly in $Z\alpha$  and in the Born approximation as
a function of the initial photon energy. For all energies
considered, the Born result exceeds the exact one by  20~\%
approximately. Therefore, the use of the Born cross section for MC
simulation would lead to the disagreement between the theory and
the experiment of 3.5 standard deviations. In other words, the
experimental results are significantly closer to predictions of
the exact theory than to those obtained in the Born approximation.

We conclude that the high-energy photon splitting in the atomic
fields is reliably observed and well investigated. This nonlinear
QED process is also studied in detail theoretically. At present,
the experiment and the theory are consistent within the achieved
experimental accuracy.

\appendix
\renewcommand{\thesection}{\appendixname\hspace{0.5em}\Alph{section}}
\renewcommand{\theequation}{\Alph{section}.\arabic{equation}}
\renewcommand{\thetable}{\Alph{section}.\arabic{table}}

\section{}
\label{AppendixWW} In this Appendix we present the result of
\cite{DeTollis64} for the amplitudes of photon-photon scattering
entering (\ref{WW:Eq:Mbar})
\begin{eqnarray}
\label{AppendixWW:Eq:M} M_{++++}&=&{8\alpha^2}\Biggl\{
 -1-\left(2+\frac{4\beta}{\zeta}\right)Q(\beta)
 -\left(2+\frac{4\gamma}{\zeta}\right)Q(\gamma)
 \nonumber\\
 &&
 -\left[\frac{2(\beta^2+\gamma^2)}{\zeta^2}+\frac8\zeta\right]
 [T(\beta)+T(\gamma)]
 -\frac4\beta\left(1+\frac2\zeta\right)I(\zeta,\beta)
 \nonumber\\
 &&
 -\frac4\gamma\left(1+\frac2\zeta\right)I(\zeta,\gamma)
 +\left[\frac{2(\beta^2+\gamma^2)}{\zeta^2}
 +\frac{16}{\zeta}+\frac4\beta+\frac4\gamma
 -\frac8{tu}\right] I(\beta,\gamma)\Biggr\}\,,
 \nonumber\\
M_{+++-}&=& {8\alpha^2}\Biggl\{
 1-4\left(\frac1\zeta+\frac1\beta+\frac1\gamma\right)
 [T(\zeta)+T(\beta)+T(\gamma)]
 \nonumber\\
 &&
 +4\left(\frac1\gamma-\frac2{st}\right)I(\zeta,\beta)
 +4\left(\frac1\beta-\frac2{su}\right)I(\zeta,\gamma)
 +4\left(\frac1\zeta-\frac2{tu}\right)I(\beta,\gamma)\Biggr\}\,,
 \nonumber\\
 M_{++--}&=& {8\alpha^2}\Biggl\{
 1-\frac8{st} I(\zeta,\beta)-\frac8{su} I(\zeta,\gamma)
 -\frac8{tu} I(\beta,\gamma)\Biggr\}\,,\nonumber\\
  M_{+-+-}&=&M_{++++}(\zeta\leftrightarrow\gamma)\,, \quad
M_{+--+}=M_{++++}(\zeta\leftrightarrow\beta)\,.
\end{eqnarray}
Here
\begin{eqnarray}
&&Q(x)=\sqrt{1+\frac4{x-i0}}\,
 \textrm{arcsinh}\left(\frac{\sqrt{x-i0}}2\right)-1\,,
\quad T(x)=\textrm{arcsinh}^2\left(\frac{\sqrt{x-i0}}2\right)\,, \nonumber\\
&&I(x,y)=\frac14\int\limits_0^1 d\tau \frac{
 \ln[1+x\tau(1-\tau)-i0]+\ln[1+y\tau(1-\tau)-i0]
 }{\tau (1-\tau)+(x+y)/xy}\nonumber\\
&& \zeta=-\frac{2k_2k_3}{m^2}=-\frac{k_{2\perp}^2}{m^2x(1-x)} \,,\
   \beta=\frac{2k_1k_2}{m^2}=\frac{k_{2\perp}^2}{m^2x}\,,\
   \gamma=\frac{2k_1k_3}{m^2}=\frac{k_{2\perp}^2}{m^2(1-x)}\, .
\end{eqnarray}
In the expressions for $\zeta$, $\beta$, and $\gamma$ we took into account the
smallness of $\theta_{2,3}$ and the relation
$\bm{k}_{3\perp}=-\bm{k}_{2\perp}$, which is valid in the region of
applicability of the WW approximation.

The asymptotics of (\ref{AppendixWW:Eq:M}) at $|\zeta|,\beta,\gamma\ll 1$ reads
\begin{eqnarray}
\label{AppendixWW:Eq:MA}
 M_{++++}&\approx&\frac{11\alpha^2}{45}\zeta^2\,,\quad
 M_{+--+}\approx\frac{11\alpha^2}{45}\beta^2\,,\quad
 M_{+-+-}\approx\frac{11\alpha^2}{45}\gamma^2\,,\nonumber\\
 M_{++--}&\approx&-\frac{\alpha^2}{15}(\zeta^2+\beta^2+\gamma^2)\,,\quad
 M_{+++-}\approx0\,.
\end{eqnarray}
At $|\zeta|,\beta,\gamma\gg 1$ one has
\begin{eqnarray}
\label{AppendixWW:Eq:MA1}
 M_{++++}&=&
 8\alpha^2\left\{
 -1+(1-2x) \ln[1/x-1]-(x^2-x+1/2)(\ln^2[1/x-1]+\pi^2)
 \right\}\,,\nonumber\\
 M_{+--+}&=&M_{++++}(x\to 1/(1-x+i0))\,,\quad
 M_{+-+-}=M_{++++}(x\to 1/(x+i0))\,,\nonumber\\
 M_{++--}&=&M_{+++-}=8\alpha^2\,.
\end{eqnarray}

\section{}\label{AppShima}
In this Appendix we present the functions $x_i$ and $y_i$ which
appear in the result for the Born cross section
(\ref{Born:Eq:Shima}) obtained in \cite{Sh} and later rederived in
\cite{CTP}.
\begin{eqnarray}
x_1&=&\zeta\left[
 \left(
  {2\zeta\over\beta}+{2\gamma-\zeta\over\beta+\gamma}
  -{2\zeta\beta\over(\beta+\gamma)^2}
  \right){\omega_1\over m}
 +\gamma\left(
  {2\over\beta}+{1\over\beta+\gamma}
  \right){\omega_2\over m}
\right.
\nonumber\\
&& \left.
 -{\beta\over\beta+\gamma}{\omega_3\over m}
\right]
 [A(\zeta)-A(\delta)]
-\left[ \gamma\left(
  {\zeta\gamma\over\beta^2}+{\zeta-\gamma-2\over2\beta}
  -{1\over\beta+\gamma}
  \right){\omega_2\over m}
\right.
\nonumber\\
&& \left. +\zeta\left(
  {\zeta\gamma\over\beta^2}-{\zeta-\gamma\over2\beta}
  -{1\over\zeta}-{1\over\beta}-{1\over\gamma}+{1\over\beta+\gamma}
  +{2\beta\over(\beta+\gamma)^2}
  \right){\omega_1\over m}
\right.
\nonumber\\
&& \left.
 -\beta\left(
  {\zeta-\gamma\over2\beta}+{\beta\over\gamma(\beta+\gamma)}
  \right){\omega_3\over m}
\right] B(\zeta) +\left[
 {\zeta(\beta-\gamma)\over \beta\gamma}{\omega_1\over m}
 +{\beta\over \gamma}{\omega_3\over m}
\right] B(\beta)
\nonumber\\
&& +\zeta\left[
 \left(
  {\zeta\gamma\over\beta^2}-{\zeta-\gamma\over2\beta}
  -{1\over\beta}-{1\over\gamma}-{1\over\zeta+\beta}+{1\over\beta+\gamma}
  +{2\beta\over(\beta+\gamma)^2}
  \right){\omega_1\over m}
\right.
\nonumber\\
&& \left.
 -\beta\left(
  {1\over2\beta}+{\beta\delta\over
  \zeta\gamma(\zeta+\beta)(\beta+\gamma)}
  \right){\omega_3\over m}
\right] B(\delta) +{2\zeta\gamma \over\beta+\gamma}{\omega_1\over m}
\nonumber\\
&& -2\left[
 {\zeta(\zeta+\gamma)\over \beta\gamma}{\omega_1\over m}
 +{(\zeta-\beta)(\zeta+\gamma)\over \zeta\beta}{\omega_2\over
  m}
 +\left({\beta\over\zeta}+{\zeta\over\gamma}\right){\omega_3\over m}
\right] S(\zeta,\beta,\gamma)
\nonumber\\
&& -\zeta\left[
 \left({2\zeta\gamma\over\beta^2}+{\gamma-\zeta-6\over\beta}
  -{4\over\zeta}+{2\over\gamma}\right){\omega_1\over
  m}
 -\left(1-{2\beta\over\zeta\gamma}\right){\omega_3\over m}
\right] S(\beta,\zeta,\gamma)
\nonumber\\
&& +(\zeta\leftrightarrow\gamma,
\omega_1\leftrightarrow\omega_2,\omega_3\rightarrow-\omega_3)\,,
\nonumber\\
x_4&=& \left[ {\zeta\over\gamma}{\omega_1\over
 m}-{\gamma(\zeta-\beta)\over \zeta\beta}{\omega_2\over m}
 \right][B(\zeta)+B(\beta)+B(\gamma)-B(\delta)]
+{2(\zeta+\gamma)\over\zeta}{\omega_2\over m}
 S(\zeta,\beta,\gamma)
\nonumber\\
&&-{2\over\beta} \left[
 (\zeta+\beta){\omega_1\over m}
 +(\beta+\gamma){\omega_2\over m}
 \right] S(\beta,\zeta,\gamma)-(\beta\leftrightarrow\gamma,\,
 \omega_2\leftrightarrow\omega_3)\,,
\nonumber\\
x_2&=&x_1(\beta\leftrightarrow\gamma,\,
 \omega_2\leftrightarrow\omega_3)\,,
\quad x_3=x_1(\zeta\leftrightarrow\beta,\,
 \omega_3\leftrightarrow-\omega_1)
\, .
\\
y_1&=&-2\left({1\over\zeta}+{\zeta+\gamma-4\over\beta\gamma}
 \right) S(\zeta,\beta,\gamma)
+\left({\zeta\gamma\over \beta^2}+{\zeta-5\over\beta}+{4-2\zeta\over
 \zeta\gamma}\right) S(\beta,\zeta,\gamma)
\nonumber\\
&& -\zeta\left({2\over\beta}-{1\over\beta+\gamma}\right)
 [A(\zeta)-A(\delta)]
+\left({\zeta\gamma\over \beta^2}+{\zeta+\gamma-6\over
 2\beta}-{\beta\over \gamma(\beta+\gamma)}\right) B(\zeta)
\nonumber\\
 &&
-\left({\zeta\gamma\over 2\beta^2}+{\zeta+\gamma-6\over
 4\beta}-{\beta\over \gamma(\beta+\gamma)}\right) B(\delta)
-\left({1\over\zeta}+{1\over 2\beta}\right) B(\beta)
+(\zeta\leftrightarrow\gamma)
 \,,
\nonumber\\
y_4&=&2-({1\over\zeta}+{1\over\beta}+{1\over\gamma})
 [B(\zeta)+B(\beta)+B(\gamma)-B(\delta)]
\nonumber\\
&& -{2\over\zeta}\left(1-{4\zeta\over\beta\gamma}\right) S(\zeta,\beta,\gamma)
-{2\over\beta}\left(1-{4\beta\over\zeta\gamma}\right) S(\beta,\zeta,\gamma)
-{2\over\gamma}\left(1-{4\gamma\over\zeta\beta}\right) S(\gamma,\zeta,\beta)\,
,
\nonumber\\
y_2&=&y_1(\zeta\leftrightarrow\beta) \,, \quad
y_3=y_1(\gamma\leftrightarrow\beta)\,.
\end{eqnarray}
Here $\delta=\zeta+\beta+\gamma$ and the functions $A$, $B$, and $S$ are
defined as
\begin{eqnarray}
&&A(u)=\int\limits_0^1 dx\, \ln[1-i0+u\, x(1-x)]\,,\quad
B(u)=\int\limits_0^1{dx\over x(1-x)} \ln[1-i0+u\, x(1-x)]\,,\nonumber
\\
&&S(u,v,w)=\frac12\int\limits_0^1 \frac{ dx\,v\,w}{u-v\,w\, x(1-x)}
\ln\left\{\frac{[1-i0+v\, x(1-x)][1-i0+w\,
x(1-x)]}{1+(u+v+w) x(1-x)}\right\}\,.\nonumber\\
\end{eqnarray}
In fact, the functions $A(u)$ and $B(u)$ are expressed via elementary functions
and $S(u,v,w)$ is expressed in terms of dilogarithmic functions \cite{Sh}.

\section{}
\label{AppendixB}

In this Appendix we present some results obtained in \cite{JMO80}
for the total ($\sigma$) and differential ($d\sigma/dx$) cross
sections of photon splitting in a screened and  Coulomb
potentials. The form factor used in the calculations was taken
from \cite{FF79}.

\begin{table}[h]
  \centering
 \caption{The total Born cross section $\sigma$ in units of $Z^2 \mathrm{(nb)}$ for
  a Coulomb potential (row $Z=0$) and screened potential as a function of
  $\omega_1$ and $Z$. The dependence on $Z$ is due to the atomic form factor.}
  \vspace{0.5cm}
\begin{tabular}{|l|c|c|c|c|c|c|c|c|} \hline
 $  \quad  \omega_1/m $&$1$&$3   $&$ 10  $&$ 30  $&$ 100 $&$ 300 $&$ 1000$&$3000 $\\
 $  Z  $&$     $&$     $&$     $&$     $&$     $&$     $&$     $&$     $\\ \hline
 $   0  $&$ 1.436\times 10^{-3}
                    $&$0.795$&$12.10$&$31.47$&$56.5 $&$80.6 $&$107.3$&$131.8$\\
 $  1       $&$ 1.436\times 10^{-3}
                    $&$0.795$&$12.10$&$31.47$&$56.4 $&$79.0 $&$97.9 $&$107.6$\\
 $  10      $&$ 1.431\times 10^{-3}
                    $&$0.795$&$12.08$&$31.13$&$54.1 $&$72.0 $&$84.5 $&$90.9 $\\
 $  20      $&$ 1.423\times 10^{-3}
                    $&$0.794$&$12.03$&$30.83$&$52.98$&$69.58$&$81.18$&$87.2 $\\
 $  50      $&$ 1.40\times 10^{-3}
                    $&$0.790$&$11.92$&$30.32$&$51.32$&$66.32$&$75.72$&$81.0 $\\
 $  68      $&$ 1.39\times 10^{-3}
                    $&$0.787$&$11.85$&$30.04$&$50.54$&$64.92$&$74.22$&$78.09$\\
 $  92      $&$ 1.378\times 10^{-3}
                    $&$0.782$&$11.78$&$29.71$&$49.74$&$63.33$&$72.07$&$75.50$\\
 \hline
    \end{tabular}
 \label{AppendixB:Table1}
\end{table}

\begin{table}[h]
  \centering
 \caption{The Born cross section $d\sigma/dx$ in units of $Z^2\mathrm{(b)}$ as a
 function of $x=\omega_2/\omega_1$
 for  different $\omega_1$. The column $Z=0$ corresponds to
 the a Coulomb potential.}
\vspace{0.5cm}
 \begin{tabular}{|l|c|c|c|c|c|}
\hline
 \multicolumn{1}{|r|}{$Z$}&$0   $&$20   $&$ 50  $&$ 82  $&$ 92  $\\
 $    x $&$    $&$     $&$     $&$     $&$     $\\
\hline\hline \multicolumn{6}{|c|}{$\omega_1=m$}
\\
 \hline $0.00993$&$   1.78\times 10^{-15}
$&$ 1.59\times 10^{-15} $&$ 1.50\times 10^{-15} $&$   1.44\times 10^{-15} $&$
1.42 \times 10^{-15}
$\\
$0.0508 $&$   1.36\times 10^{-13} $&$   1.27\times 10^{-13} $&$ 1.22 \times
10^{-13} $&$   1.19\times 10^{-13} $&$   1.17\times 10^{-13}
$\\
$0.119 $&$   9.09\times 10^{-13} $&$   8.84 \times 10^{-13} $&$   8.63\times
10^{-13} $&$   8.47\times 10^{-13} $&$   8.41 \times 10^{-13}
$\\
$0.204 $&$   2.42 \times 10^{-12} $&$   2.39\times 10^{-12} $&$   2.34\times
10^{-12} $&$   2.31\times 10^{-12} $&$   2.30\times 10^{-12}
$\\
$0.296 $&$   3.88\times 10^{-12} $&$   3.85\times 10^{-12} $&$   3.79\times
10^{-12} $&$   3.74\times 10^{-12} $&$   3.73\times 10^{-12}
$\\
$0.381 $&$   4.78 \times 10^{-12} $&$   4.74 \times 10^{-12} $&$   4.68 \times
10^{-12} $&$   4.63 \times 10^{-12} $&$   4.61 \times 10^{-12}
$\\
$0.449 $&$   5.09\times 10^{-12} $&$   5.06 \times 10^{-12} $&$   4.99 \times
10^{-12} $&$   4.93\times 10^{-12} $&$   4.92 \times 10^{-12}
$\\
$0.49 $&$   5.08\times 10^{-12} $&$   5.05\times 10^{-12} $&$   4.98\times
10^{-12} $&$   4.93 \times 10^{-12} $&$   4.91\times 10^{-12}$ \\
\hline\hline \multicolumn{6}{|c|}{$\omega_1=10m$}
\\
\hline $0.00993 $&$   1.84\times 10^{-9} $&$   1.79\times 10^{-9} $&$
1.76\times 10^{-9} $&$   1.73\times 10^{-9} $&$   1.72\times 10^{-9}
$\\
$0.0508 $&$   2.51 \times 10^{-8} $&$   2.49\times 10^{-8} $&$   2.46\times
10^{-8} $&$   2.44 \times 10^{-8} $&$   2.43\times 10^{-8}
$\\
$0.119 $&$   2.76\times 10^{-8} $&$   2.75\times 10^{-8} $&$   2.72\times
10^{-8} $&$   2.70\times 10^{-8} $&$   2.69\times 10^{-8}
$\\
$0.204 $&$   2.64\times 10^{-8} $&$   2.63\times 10^{-8} $&$   2.60\times
10^{-8} $&$   2.58\times 10^{-8} $&$   2.57\times 10^{-8}
$\\
$0.296 $&$   2.51\times 10^{-8} $&$   2.49\times 10^{-8} $&$   2.47\times
10^{-8} $&$   2.45\times 10^{-8} $&$   2.44\times 10^{-8}
$\\
$0.381 $&$   2.42\times 10^{-8} $&$   2.41\times 10^{-8} $&$   2.39\times
10^{-8} $&$   2.37\times 10^{-8} $&$   2.36\times 10^{-8}
$\\
$0.449 $&$   2.39\times 10^{-8} $&$   2.37\times 10^{-8} $&$   2.35\times
10^{-8} $&$   2.33\times 10^{-8} $&$   2.32\times 10^{-8}
$\\
$0.49 $&$   2.38\times 10^{-8} $&$   2.37\times 10^{-8} $&$   2.34 \times
10^{-8} $&$   2.32\times 10^{-8} $&$   2.32 \times 10^{-8}
$\\
\hline\hline \multicolumn{6}{|c|}{$\omega_1=100m$}
\\
\hline $0.00993 $&$   1.65\times 10^{-7} $&$   1.56\times 10^{-7} $&$
1.52\times 10^{-7} $&$   1.48\times 10^{-7} $&$   1.47\times 10^{-7}
$\\
$0.0508 $&$   1.49\times 10^{-7} $&$   1.40 \times 10^{-7} $&$   1.36\times
10^{-7} $&$   1.33\times 10^{-7} $&$   1.32\times 10^{-7}
$\\
$0.119 $&$   1.28\times 10^{-7} $&$   1.20\times 10^{-7} $&$   1.16\times
10^{-7} $&$   1.13\times 10^{-7} $&$   1.13\times 10^{-7}
$\\
$0.204 $&$   1.11\times 10^{-7} $&$   1.04\times 10^{-7} $&$   1.01\times
10^{-7} $&$   9.83\times 10^{-8} $&$   9.76\times 10^{-8}
$\\
$0.296 $&$   1.01\times 10^{-7} $&$   9.39\times 10^{-8} $&$   9.09\times
10^{-8} $&$   8.86\times 10^{-8} $&$   8.80 \times 10^{-8}
$\\
$0.381 $&$   9.48 \times 10^{-8} $&$   8.85\times 10^{-8} $&$   8.56\times
10^{-8} $&$   8.35\times 10^{-8} $&$   8.29\times 10^{-8}
$\\
$0.449 $&$   9.26\times 10^{-8} $&$   8.64 \times 10^{-8} $&$   8.36\times
10^{-8} $&$   8.15\times 10^{-8} $&$   8.09\times 10^{-8}
$\\
$0.49 $&$   9.21\times 10^{-8} $&$   8.59\times 10^{-8} $&$   8.31 \times
10^{-8} $&$   8.10\times 10^{-8} $&$   8.05\times 10^{-8} $
\\
\hline\hline \multicolumn{6}{|c|}{$\omega_1=1000m$}
\\
\hline $0.00993 $&$   3.73\times 10^{-7} $&$   2.96\times 10^{-7} $&$
2.77\times 10^{-7} $&$   2.69\times 10^{-7} $&$   2.65\times 10^{-7}
$\\
$0.0508 $&$   3.01 \times 10^{-7} $&$   2.33\times 10^{-7} $&$   2.18\times
10^{-7} $&$   2.11\times 10^{-7} $&$   2.08 \times 10^{-7}
$\\
$0.119 $&$   2.44 \times 10^{-7} $&$   1.85\times 10^{-7} $&$   1.73 \times
10^{-7} $&$   1.67\times 10^{-7} $&$   1.65\times 10^{-7}
$\\
$0.204 $&$   2.06 \times 10^{-7} $&$   1.54\times 10^{-7} $&$   1.44 \times
10^{-7} $&$   1.39\times 10^{-7} $&$   1.37\times 10^{-7}
$\\
$0.296 $&$   1.83\times 10^{-7} $&$   1.36\times 10^{-7} $&$   1.27\times
10^{-7} $&$   1.22\times 10^{-7} $&$   1.21\times 10^{-7}
$\\
$0.381 $&$   1.71\times 10^{-7} $&$   1.27\times 10^{-7} $&$   1.18\times
10^{-7} $&$   1.14\times 10^{-7} $&$   1.12\times 10^{-7}
$\\
$0.449 $&$   1.67\times 10^{-7} $&$   1.23 \times 10^{-7} $&$   1.15\times
10^{-7} $&$   1.11\times 10^{-7} $&$   1.09\times 10^{-7}
$\\
$0.49 $&$   1.66 \times 10^{-7} $&$   1.23\times 10^{-7} $&$   1.14\times
10^{-7} $&$   1.10\times 10^{-7} $&$   1.08 \times 10^{-7}
$\\
\hline
\end{tabular}
 \label{AppendixB:Table2}
\end{table}

\section{}
\label{AppendixC}

Here we present the photon splitting amplitudes exact in the parameter
$Z\alpha$ and valid in the high-energy zero-mass limit
($\omega_2\theta_2,\,\omega_3\theta_3\gg m$ and $\theta_{2,3}\ll 1$)
\cite{LMS2}. They read
\begin{eqnarray}\label{AppendixC:FINAL1} &&
M=\frac{4e^3Z\alpha}{\pi\omega_1\omega_2\omega_3{\Delta}^2}
\int\limits_{-1}^{1}\frac{dy\,\,\mbox{sign}\,y}{1-y^2}\,\left[\mbox{Re}
\left(\frac{1+y}{1-y}\right)^{iZ\alpha}\right]\, R\, ; \nonumber\\
&&
 R_{+--}=\frac{q^2}{(\bm{e}^*\cdot\bm{\theta}_3)(\bm{e}^*\cdot\bm{\theta}_{23})}
\int\limits_{0}^{\omega_2}\!
d\varepsilon\,\kappa_2\,\bm{e}^*\cdot(\kappa_2\bm{\theta}_2-\bm{\Delta})
\,\frac{\vartheta(\bm{r}^2-q^2)}{(\bm{e}^*\cdot\bm{r})^2}\,+\,
{{\omega_2\leftrightarrow\omega_3}
\choose{\bm{\theta}_2\leftrightarrow\bm{\theta}_3}}\,,
\nonumber\\
&& R_{+++}=\int\limits_{0}^{\omega_2}\! d\varepsilon \,
\Biggl[\frac{\omega_3\varepsilon\kappa_2}{\omega_2}\left(
\frac{s}{t}-1\right)\left(\frac{\omega_3}{(\bm{e}\cdot\bm{u})}-
\frac{8(\bm{e}^*\cdot\bm{\theta}_3)(\kappa_2^2+\kappa_3^2)}{s-t-
4(\bm{e}^*\cdot\bm{r})(\bm{e}\cdot\bm{u})}\right)
\nonumber  \\
&&
+\vartheta(q^2-{r}^2)\frac{(\kappa_2^2+\kappa_3^2)\kappa_2(1+1/y)(\bm{e}^*\cdot\bm{\theta}_{2})(\bm{e}\cdot\bm{\Delta})}
{(\bm{e}\cdot\bm{\theta}_{23})[\varepsilon(\bm{e}^*\cdot\bm{r})(\bm{e}\cdot\bm{\theta}_{23})-\kappa_2(1+1/y)(\bm{e}\cdot\bm{\Delta})(\bm{e}^*\cdot\bm{\theta}_2)]}
\Biggr]\,+ \,{{\omega_2\leftrightarrow\omega_3}
\choose{\bm{\theta}_2\leftrightarrow\bm{\theta}_3}}\,,
\nonumber\\
&& R_{++-}=\int\limits_{0}^{\omega_2}\! d\varepsilon \,
\Biggl[\frac{\omega_3\kappa_2\kappa_3}{\omega_1}
\left(\frac{is_1}{t_1}-1\right)\left(
\frac{\kappa_2-\varepsilon}{(\bm{e}^*\cdot\bm{u}_1)}+\frac{8(\bm{e}\cdot\bm{\theta}_{23})(\kappa_2^2+\varepsilon^2)}
{s_1+it_1-4(\bm{e}\cdot\bm{r})(\bm{e}^*\cdot\bm{u}_1)}\right)
\nonumber\\
&&
+\vartheta(q^2-{r}^2)\frac{(\kappa_2^2+\varepsilon^2)\kappa_2(1+1/y)(\bm{e}\cdot\bm{\theta}_{2})(\bm{e}^*\cdot\bm{\Delta})}
{(\bm{e}^*\cdot\bm{\theta}_{3})[\kappa_3(\bm{e}\cdot\bm{r})(\bm{e}^*\cdot\bm{\theta}_{3})-\kappa_2(1+1/y)(\bm{e}^*\cdot\bm{\Delta})(\bm{e}\cdot\bm{\theta}_2)]}
\Biggr]
\nonumber\\
&& -\omega_2\int\limits_{-\omega_3}^{0}\! d\varepsilon \,
\kappa_3\Biggl[\frac{(\bm{e}\cdot\bm{\Delta})}{\omega_1(\bm{e}^*\cdot\bm{\Delta})}
\left(\frac{is_1}{t_1}-1\right)
\left(\frac{\varepsilon\kappa_3+\kappa_2^2}{(\bm{e}\cdot\bm{u}_1)}+\frac{8(\bm{e}^*\cdot\bm{\theta}_{23})\kappa_3(\kappa_2^2+\varepsilon^2)}
{s_1-it_1-4(\bm{e}^*\cdot\bm{r}_1)(\bm{e}\cdot\bm{u}_1)}\right)
\nonumber  \\
&& +\frac{(\bm{e}\cdot\bm{\Delta})}{\omega_3(\bm{e}^*\cdot\bm{\Delta})}
\left(\frac{s_2}{t_2}-1\right)
\left(\frac{\kappa_2\kappa_3-\varepsilon^2}{(\bm{e}\cdot\bm{u}_2)}+\frac{8(\bm{e}^*\cdot\bm{\theta}_{2})\kappa_3(\kappa_2^2+\varepsilon^2)}
{s_2+t_2-4(\bm{e}^*\cdot\bm{r}_1)(\bm{e}\cdot\bm{u}_2)}\right)
\nonumber  \\
&&+
\vartheta({r}^2_1-q^2)\frac{q^2(\kappa_2^2+\varepsilon^2)}{2\omega_2(\bm{e}^*\cdot\bm{r}_1)}
\Biggl(\frac{1}{\kappa_2(\bm{e}^*\cdot\bm{r}_1)
(\bm{e}\cdot\bm{\theta}_{2})-\kappa_3(1/y-1)(\bm{e}\cdot\bm{\Delta})
(\bm{e}^*\cdot\bm{\theta}_3)}
\nonumber  \\
&&
+\frac{1}{\varepsilon(\bm{e}^*\cdot\bm{r}_1)(\bm{e}\cdot\bm{\theta}_{23})-\kappa_3(1/y-1)(\bm{e}\cdot\bm{\Delta})(\bm{e}^*\cdot\bm{\theta}_3)}
\Biggr)\Biggr]\, ,
\\
{}\nonumber\\
&& \label{AppendixC:FINALadd}
\bm{u}=\bm{\Delta_\perp}\left(\frac{1}{y}-1+\frac{2\kappa_2}{\omega_2}\right)\,
,\quad
\bm{u}_1=\bm{\Delta_\perp}\left(\frac{1}{y}-1+\frac{2\kappa_2}{\omega_1}\right)\,
,\quad
\bm{u}_2=\bm{\Delta_\perp}\left(\frac{1}{y}-1-\frac{2\varepsilon}{\omega_3}\right)\, ,\,\nonumber \\
&& q^2={\Delta_\perp}^2(1/y^2-1)\,
,\quad\bm{r}=\bm{\Delta_\perp}(1/y-1)+2\kappa_2\bm{\theta}_2\, ,\quad
\bm{r}_1=\bm{\Delta_\perp}(1/y+1)-2\kappa_3\bm{\theta}_3\, ,\nonumber\\
&& \bm{\theta}_{23}=\bm{\theta}_{2}-\bm{\theta}_{3}\,, \quad
\kappa_2=\omega_2-\varepsilon\,,\quad \kappa_3=\omega_3+\varepsilon\,,\quad
s={u}^2+q^2+\frac{4\omega_1\omega_3\kappa_2\varepsilon}{\omega_2^2}{\theta}_3^2\,
,
\nonumber\\
&&
s_1={u}_1^2+q^2-\frac{4\omega_2\omega_3\kappa_2\kappa_3}{\omega_1^2}{\theta}_{23}^2\,-i0\,
,\quad
s_2={u}_2^2+q^2-\frac{4\omega_1\omega_2\kappa_3\varepsilon}{\omega_3^2}{\theta}_2^2\,
\nonumber\\
&& t=\sqrt{s^2-4q^2{u}^2}\,,\quad t_1=\sqrt{4q^2{u}_1^2-s_1^2}\,, \quad
t_2=\sqrt{s_2^2-4q^2{u}_2^2}\,.
\end{eqnarray}

In the Born approximation (linear in $Z\alpha$ term of
(\ref{AppendixC:FINAL1})) all integrals can be taken with the result
\begin{eqnarray}\label{AppendixC:Born} &&
M_{+--}=\frac{2Z\alpha e^3
[(\bm{e}^*\cdot\bm{\theta}_2)(\bm{e}\cdot\bm{\theta}_3)
-(\bm{e}\cdot\bm{\theta}_2)(\bm{e}^*\cdot\bm{\theta}_3)]}{\pi{\Delta}^2(\bm{e}^*\cdot\bm{\theta}_2)
(\bm{e}^*\cdot\bm{\theta}_3) (\bm{e}^*\cdot\bm{\theta}_{23})}\,,\nonumber\\
&& M_{+++}=\frac{2(Z\alpha) e^3\omega_1}{\pi{\Delta}^2
(\bm{e}\cdot\bm{\theta}_{23})^2\omega_2\omega_3}\Biggl\{(\bm{e}\cdot\bm{\Delta})\Biggl[1+\frac{(\bm{e}\cdot\bm{\theta}_2)+(\bm{e}\cdot\bm{\theta}_3)}
{(\bm{e}\cdot\bm{\theta}_{23})}\ln({a_2\over a_3})
\nonumber\\
&&
+\frac{(\bm{e}\cdot\bm{\theta}_2)^2+(\bm{e}\cdot\bm{\theta}_3)^2}{(\bm{e}\cdot\bm{\theta}_{23})^2}\left({\pi^2\over
6} +{1\over 2} \ln^2({a_2\over
a_3})+\mbox{Li}_2(1-a_2)+\mbox{Li}_2(1-a_3)\right)\Biggr]\nonumber\\
&& +{1\over
(\bm{e}\cdot\bm{\Delta})}\left[\omega_3^2(\bm{e}\cdot\bm{\theta}_3)^2\frac{a_2}{1-a_2}\left(
1+{a_2\,\ln(a_2)\over
1-a_2}\right)+\omega_2^2(\bm{e}\cdot\bm{\theta}_2)^2\frac{a_3}{1-a_3}\right.
\nonumber\\
&& \left.\times \left( 1+{a_3\,\ln(a_3)\over 1-a_3}\right)
\right]+\frac{2(\bm{e}\cdot\bm{\theta}_2)(\bm{e}\cdot\bm{\theta}_3)}{(\bm{e}\cdot\bm{\theta}_{23})}\left[\omega_3{a_2\,
\ln(a_2)\over 1-a_2}-
\omega_2{a_3\,\ln(a_3)\over 1-a_3}\right]\Biggr\}\, ,\nonumber \\
\nonumber\\
&& M_{++-}=\frac{2(Z\alpha) e^3\omega_2}{\pi{\Delta}^2
(\bm{e}^*\cdot\bm{\theta}_{3})^2\omega_1\omega_3}\Biggl\{(\bm{e}^*\cdot\bm{\Delta})\Biggl[1-
\frac{(\bm{e}^*\cdot\bm{\theta}_2)+(\bm{e}^*\cdot\bm{\theta}_{23})}{(\bm{e}^*\cdot\bm{\theta}_3)}\ln({-a_1\over
a_2})
\nonumber\\
&& +
\frac{(\bm{e}^*\cdot\bm{\theta}_2)^2+(\bm{e}^*\cdot\bm{\theta}_{23})^2}{(\bm{e}^*\cdot\bm{\theta}_3)^2}\left({\pi^2\over
6} +{1\over 2} \ln^2({-a_1\over
a_2})+\mbox{Li}_2(1-a_2)+\mbox{Li}_2(1+a_1)\right)\Biggr]\nonumber\\
&& +{1\over
(\bm{e}^*\cdot\bm{\Delta})}\left[\omega_3^2(\bm{e}^*\cdot\bm{\theta}_{23})^2\frac{a_2}{1-a_2}\left(
1+{a_2\,\ln(a_2)\over
1-a_2}\right)-\omega_1^2(\bm{e}^*\cdot\bm{\theta}_2)^2\frac{a_1}{1+a_1} \right.
\nonumber\\
&& \left.\times\left( 1-{a_1\,\ln(-a_1)\over 1+a_1}\right)
\right]+\frac{2(\bm{e}^*\cdot\bm{\theta}_2)(\bm{e}^*\cdot\bm{\theta}_{23})}{(\bm{e}^*\cdot\bm{\theta}_3)}\left[
\omega_1{a_1\,\ln(-a_1)\over 1+a_1} -\omega_3{a_2\,\ln(a_2)\over
1-a_2}\right]\Biggr\}\, ,
\end{eqnarray}
where
\begin{eqnarray}
&&a_1=\frac{{\Delta_\perp}^2}{\omega_2\omega_3\theta_{23}^2}-i0\,,\quad
a_2=\frac{{\Delta_\perp}^2}{\omega_1\omega_2\theta_2^2}\,,\quad
a_3=\frac{{\Delta_\perp}^2}{\omega_1\omega_3\theta_3^2}\,,
\quad\mathrm{Li}_2(x)=-\int\limits_0^x\frac{dt}{t}\ln(1-t)\,.\nonumber
\end{eqnarray}
Let us introduce the variable
$\bm{\varrho}=(\omega_2\bm{\theta}_2-\omega_3\bm{\theta}_3)/2$. If $\Delta\ll
\varrho$ then the asymptotics of (\ref{AppendixC:Born}) has the form
\begin{eqnarray}\label{AppendixC:Eq:zero}
M_{+--}&=&\frac{4N(\bm{e}\cdot\bm{\varrho})^3}{{\varrho}^4}
[(\bm{e}^*\cdot\bm{\Delta})(\bm{e}\cdot\bm{\varrho})
-(\bm{e}\cdot\bm{\Delta})(\bm{e}^*\cdot\bm{\varrho})]\quad,\quad
N=\frac{4Z\alpha{e^3}\omega_2\omega_3}{\pi\omega_1{\Delta}^2{\varrho}^2}\,;
\nonumber\\
 M_{+++}&=&N \Biggl[\bm{e}^*\cdot\bm{\Delta}+
2(\bm{e}\cdot\bm{\Delta})\frac{(\bm{e}^*\cdot\bm{\varrho})^2}{{\varrho}^2}\left(1+
\frac{\omega_2-\omega_3}{\omega_1}\ln\frac{\omega_3}{\omega_2}+
\frac{\omega_2^2+\omega_3^2}{2\omega_1^2}(\ln^2\frac{\omega_3}{\omega_2}+\pi^2)\right)
\Biggr]\nonumber\\
 M_{++-}&=&N
\Biggl[\bm{e}\cdot\bm{\Delta}+2(\bm{e}^*\cdot\bm{\Delta})\frac{(\bm{e}\cdot\bm{\varrho})^2}{{\varrho}^2}\left(1+
\frac{\omega_1+\omega_3}{\omega_2}(\ln\frac{\omega_3}{\omega_1}+i\pi)
\right.\nonumber\\
&& +
\left.\frac{\omega_1^2+\omega_3^2}{2\omega_2^2}(\ln^2\frac{\omega_3}{\omega_1}+
2i\pi\ln\frac{\omega_3}{\omega_1})\right) \Biggr]\, .
\end{eqnarray}

The contribution of the region $\Delta\ll \varrho$ to the total cross section
is logarithmically amplified and can be obtained also within the WW
approximation. The Coulomb corrections to the amplitudes in this region are
small compared to (\ref{AppendixC:Eq:zero}) and have the form
\begin{eqnarray}
\label{AppendixC:Eq:CCzero}
M_{+--}^{(c)}&=&-\frac{2e^3(Z\alpha)^3\omega_2\omega_3\,(\bm{e}^*\cdot\bm{\Delta})}
{\pi\omega_1(\bm{e}^*\cdot\bm{\varrho})^4}\ln^2\frac{\varrho}{\Delta}\quad ;\\
{}&&
\nonumber\\
M_{+++}^{(c)}&=&-\frac{ie^3Z\alpha\omega_2\omega_3}{2\pi\omega_1\varrho^2(\bm{e}\cdot\bm{\varrho})
(\bm{e}^*\cdot\bm{\Delta})} \int\frac{d\bm{q}}{\bm{e}\cdot(\bm{q}-\bm{\Delta})}
\nonumber\\
&& \times\left[
\mathrm{Re}\,\left(\frac{|\bm{q}+\bm{\Delta}|}{|\bm{q}-\bm{\Delta}|}\right)^{2iZ\alpha}-1\right]\,
\mathrm{sign}[(\bm{q}-\bm{\Delta})\times\bm{\varrho}]_z\quad ;\nonumber\\
M_{++-}^{(c)}&=& -\frac{e^3Z\alpha\omega_2^2\omega_3}{2\pi^2\omega_1^2
\varrho^2(\bm{e}^*\cdot\bm{\varrho})(\bm{e}\cdot\bm{\Delta})}
\int\frac{d\bm{q}}{\bm{e}^*\cdot(\bm{q}-\bm{\Delta})}
\nonumber\\
&& \times\left[
\mathrm{Re}\,\left(\frac{|\bm{q}+\bm{\Delta}|}{|\bm{q}-\bm{\Delta}|}\right)^{2iZ\alpha}-1\right]\,
\left[\ln\frac{|\bm{q}-\bm{\Delta}|}{\Delta}+
i\,\mathrm{arg}\frac{\bm{e}\cdot(\bm{q}-\bm{\Delta})}{\bm{e}\cdot\bm{\varrho}}\right]\,
.\nonumber
\end{eqnarray}
It follows from (\ref{AppendixC:Eq:CCzero}) that in this limiting case the
Coulomb correction $M_{+--}^{(c)}$ is small, while $M_{+++}^{(c)}$ and
$M_{++-}^{(c)}$ depend only on the direction of vector $\bm{\Delta}$, but not
on its module (it becomes obvious after the substitution $\bm{q}\rightarrow
\bm{q} \,\Delta$).

\end{document}